# Networking - A Statistical Physics Perspective

Chi Ho Yeung and David Saad

*The Nonlinearity and Complexity Research Group, Aston University, Birmingham B4 7ET, United Kingdom*

**Abstract**

Efficient networking has a substantial economic and societal impact in a broad range of areas including transportation systems, wired and wireless communications and a range of Internet applications. As transportation and communication networks become increasingly more complex, the ever increasing demand for congestion control, higher traffic capacity, quality of service, robustness and reduced energy consumption require new tools and methods to meet these conflicting requirements. The new methodology should serve for gaining better understanding of the properties of networking systems at the macroscopic level, as well as for the development of new principled optimization and management algorithms at the microscopic level. Methods of statistical physics seem best placed to provide new approaches as they have been developed specifically to deal with non-linear large scale systems. This paper aims at presenting an overview of tools and methods that have been developed within the statistical physics community and that can be readily applied to address the emerging problems in networking. These include diffusion processes, methods from disordered systems and polymer physics, probabilistic inference, which have direct relevance to network routing, file and frequency distribution, the exploration of network structures and vulnerability, and various other practical networking applications.

*Keywords:* Networking, statistical physics, networks, transportation, communication, random walk, disordered systems, inference.

# Contents

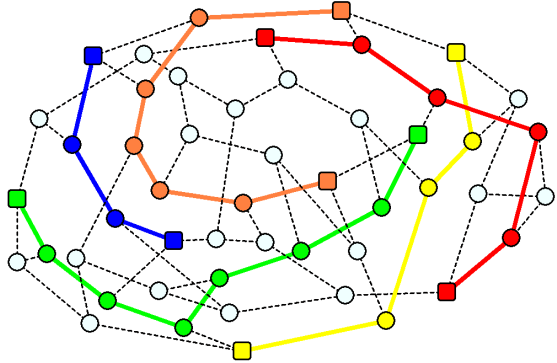

Figure 1: An example of routing configuration on a network: pair of square nodes with specific color correspond to a specific source-destination pair, circle nodes and edges with specific color correspond to the path of the corresponding source-destination pair.

## 1. Introduction

Networking encompasses a variety of tasks related to the communication of information on networks; for instance routing, frequency allocation, information spreading, distributed storage and dynamic network exploration. One particularly important aspect of networking is routing that corresponds to path selection on networks to achieve a given objective, for instance, to establish a communication path for each of the multiple source-destination pairs as shown schematically in Fig. 1, to search for a certain node on a network, or to send a message to a specific set of nodes. Networks determine the underlying topology for most networking tasks, usually correspond to a group of nodes, such as personal computers or geographic locations, connected by edges, such as transmission lines or roads.

Indeed, networking and path selection are at the heart of many communication and logistics applications. One of the most visible examples is the Internet, which connect computers and servers around the world by transmitting data through wired and wireless connections from specific sources to specific destinations [1, 2]. Due to the expanding coverage of the Internet, various overlay networks emerge whose functionality relies on effective networking. These include peer-to-peer networks (P2P) of individual users who share overlay network resources; messenger networks that link up users and allow for instant messages to be transmitted between them; and applications such as Facebook that help users establish a virtual social network [3]. From their ever increasing popularity, it is clear that routing and path selection have become an essential part in our way of life.

Other than the Internet, effective networking is essential to many daily essential applications we take for granted. For instance, transportation networks involve a large number of simultaneous path selections which constitute complex traffic dynamics [4]. Another example is sensor networks, for example fire and pollution sensors, which involve spatially distributed sensors monitoring local conditions and sending messages to a centralized base station to be communicated further [5]. Networking also represents the distribution of water in water supply network [6], communication to end-recipient in cloud computing [7], the allocation of computing powers in computer clusters [8], and even efficient coordination of military convoys movement between geographic locations [9].

Due to its wide range of applications and significance, understanding fundamental aspects and the dynamics of networking has become part of an important cross-disciplinary effort. Researchers in engineering, computer science and physics have been eager to apply their techniques for addressing various aspects of this task. For instance, engineers and computer scientists are interested in the design of practical protocols to satisfy various operational constraints such as power capacity, bandwidth and buffer size [10, 11, 12]. Researchers in management science optimize routing and scheduling for low-cost and efficient transportation [13]. The contribution of the physics community has many facets: understanding the interaction between competing communication and paths; the interplay between topology and routing dynamics; the derivation of macroscopic phenomenon such as phase transitions and how they determine the optimal operational conditions; how communication frequencies, codes and data can be optimally deployed; and the development of methods to better understand the dynamically changing topology from a minimal number of measurements.

The present article will focus on the use of methods from statistical mechanics in the study of networking. We will review basic methods and network models used in the study of traffic jams and routing, the spread of computer viruses, frequency assignment for wireless routers, efficient load balancing, resource trafficking, broadcast and multicast on networks and path optimization. We will discuss the potential algorithmic implications derived from the study of disordered systems and the use of distributed probabilistic methods for managing such systems. In addition, we will point to the potential such methods hold for the future study and control of networking systems represented via the network metaphor.

We note that the present review focuses on networking rather than network and complex systems studied in a number of review articles in the physics literature. These include reviews on structural properties [14, 15, 16], mathematical proofs on network diameters and path lengths [17], network critical phenomenon and spin models [18], dynamical processes and synchronization [16, 19, 18, 20], and community detection in networks [21].

*1.1. Practical Networking Methods*

To understand the contribution of physics to networking problems and the potential they hold for future development, we describe here a few commonly used heuristic methods employed in the Internet and practical wireless networks, which aim to solve specific networking tasks and instances.

*1.1.1. Routing*

*Table-based routing methods* is a family of techniques used for Internet routing. To use this method one first has to assign an address to each node in the network, for instance, the so-called Internet Protocol (IP) address. The next step is to compute a routing table for each individual node identifying paths to all other nodes in the network, as similar to those indicated for node 1 and 4 in Fig. 2(a). These tables can be computed to satisfy a predefined goal, for example to identify the shortest path; minimal path weight algorithms can be used [22, 23] for optimal routing. A simple example of a routing procedure is given

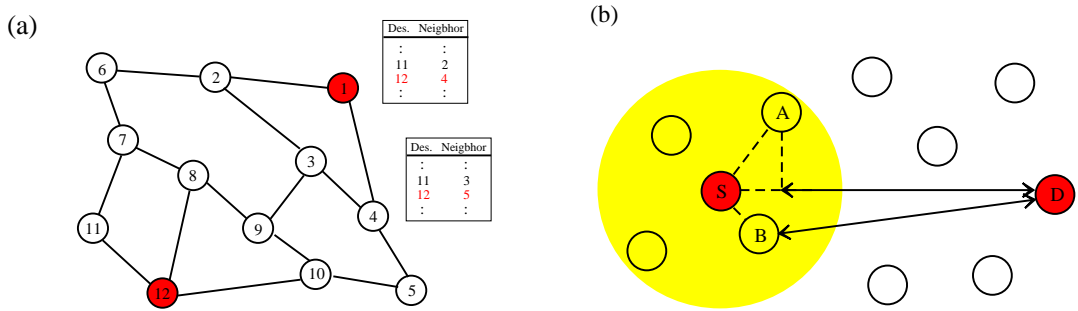

Figure 2: (a) A simple example of a table-based routing method in sending a message from node 1 to node 12. (b) An example of position-based routing algorithm in a routing message from node S to node D.

in Fig. 2(a): suppose node 1 has to send a message to node 12, it looks for the entry of node 12 in its own routing table which indicates that a message to node 12 should be sent via node 4. The message is then delivered to node 4, whose routing table states that the message to node 12 should be sent via node 5. The process continues until the message is delivered to node 12. The success of such table-based methods depend crucially on the choice of paths indicated by the routing tables, which are usually pre-calculated and are non-adaptive to changes in topology and traffic conditions, hence may result in sub-optimal paths when certain routers break down or when traffic is congested. Attempts to resolve these issues include the calculation of multiple paths for each source-destination pair [24, 25], or the use of a *reactive* approach to repeatedly re-calculate the routing tables.

Next, we describe the *position-based methods* [26, 27, 28, 29], used in wireless network where the positions of nodes are known, for example, by GPS (global positioning system). In general, these methods work for networks with frequent topology changes and switches between individual active and inactive states. A simple example is shown in Fig. 2(b): suppose node $S$ has to send a message to node $D$, it forwards the message to the most suitable node in its neighborhood limited by its own transmission range. There are various approaches for determining the must suitable node, for example, $S$ may forward the message to node $A$ since $A$ has the shortest distance to $D$ [26] projected along the line connecting $S$ and $D$, or to $B$, since $B$ has the shortest Euclidean distance to $D$ [27]. In these greedy forwarding scenarios, problems such as the absence of a suitable neighbor or an infinite looping of messages may arrive. Attempts to resolve these issues include the introduction of recovery modes, for example reset routing around the perimeter of the trapped region, defined as the region surrounded by a loopy transition of messages [28].

### 1.1.2. Frequency allocation

Other networking tasks that are currently solved heuristically are, for instance, frequency or channel allocation in cellular networks and radio broadcasting systems. The goal is to suppress interference between neighboring areas where similar broadcasting radio frequencies are used. For a small number of radio or television channels, one may manually assign different frequencies to neighboring base stations. However, the problem becomes more complicated and infeasible to carry out manually in communication networks with a large number of wireless devices, each relying on an individual communication channel. The problem can be mapped onto a graph coloring problem where no adjacent (interacting) nodes share the same color (see Section 4.1.1 for details). Heuristic methods of frequency allocation are proposed in Refs. [30] and [31] in accordance with the graph coloring optimization problem. In these methods, channels are consecutively assigned to users according to criteria such as minimization of the number of potentially conflicting neighbors or the decrease in overall channel availability, until all users are assigned frequencies such that no neighboring assignments are in conflict. In addition to assignment on fixed graphs, methods have also been devised to tackle the problem of adaptive networks of constantly changing topologies [32].

Another aspect of the same problem is in the distributive storage of file segments such that files could be reassembled efficiently upon request. One can map this problem onto a diversification of color assignment, where each color corresponds to one file segment, such that for each node a sufficient number of heterogeneous colors (i.e. file segments) are found in the immediate neighborhood for file reconstruction. Details are found in Section 4.1.2 and Ref. [33]. Various heuristic algorithms have been proposed to optimize color diversify in graphs of different topology [33, 34]. For instance, in tree networks one can adopt the method suggested in Ref. [33] to start the color assignment by first coloring the root node and its nearest neighborhood with the set of all required colors. The process then continues in a manner similar to epidemic spreading, by coloring nearest empty node to satisfy nodes which are closest to the root that do not have all the necessary colors. Finally, one obtains a colored tree which corresponds to an assignment of file segments where every node can retrieve a complete set of file segments from its immediate neighborhood.

*1.1.3. Dynamical network discovery*

A technology called route analytics can be employed to monitor dynamical changes in topology or latency of a router network. In this approach, an analysis appliance is introduced to monitor relations between routers and obtain dynamic information of the network. When routers exchange information to decide on the routing of particular packets, the appliance acts as a passive router and passively receive all the information (i.e., it does not participate in forwarding the packets). The information obtained from the appliance provides a real-time overview of the state of the network . Other than direct analysis of routing information, individual link latency may also be inferred from end-to-end information, a process which is termed network tomography. It involves solving an under-determined system of linear equations (details are described in Section 5.3). Progress has been made within the statistical physics community on identifying the conditions for which information on link latency can be accurately retrieved as well as in devising methods for solving individual instances; traditional techniques rely on linear programming techniques to solve individual instances [35, 36].

*1.2. Statistical Physics and Networking*

As we can see from the previous subsection, there are already a variety of ingenious methods for solving specific networking tasks. However, such algorithmic studies often offer a heuristic sub-optimal microscopic solutions for solving real instances, leaving the macroscopic behaviors such as phase transitions and self-organization less understood. Statistical physics offers insight into the macroscopic behavior of networking from basic definition of the microscopic interactions, through the use of established techniques such as spin glass theory [37, 38, 39, 40], which take into account the disorder induced by the network structure or communicating nodes. It also facilitates the development of principled algorithms for solving efficiently some of the more difficult networking tasks.

In the past decade statistical mechanics has been applied to a variety of problems which have origins outside the traditional realm of physics. New insight and understanding that is inaccessible via traditional techniques have been gained, which have had a substantial impact in the respective fields and have led to new research directions and the development of new methodology. Among all these research activities are diverse problems such as combinatorial optimization, epidemic spreading and information theory, which facilitate the understanding of the appropriate operational regimes and their limitations in the networking context.

In this article, we will review the applications of physics techniques of both proven and potential impact on networking tasks. In Section 2 we will review basic network properties and some of the most commonly used network models that are essential in the study of networking problems. In Section 3, we will review the master equation and Markov chain techniques employed to describe probabilistic flow in networks [41], which are relevant to network search and epidemic spreading in computer networks. In Section 4, we will review the use of spin glass theory of disordered systems in addressing networking problems such as wireless frequency allocation, resource allocation and path optimization. In Section 5, we will review cross-disciplinary links identified between statistical physics and established concepts in probability and statistics such as Bayesian methods, maximum likelihood, variational approaches and message passing techniques [39], which have a direct implication in networking applications. Finally we will mention future directions in the potential uses of the methods mentioned in the review to networking.

## 2. Networks Essentials

As networks constitute the underlying topology used in most networking tasks, we will describe the terminology and notation used in the article in Section 2.1, essential features and measures in Section 2.2, and the network models studied or introduced by physicists in Section 2.3. Since network features and models have been included in a number of review articles [14, 15, 16], we recommend to readers who are familiar with complex networks models and terminology to skip Sections 2.2 and 2.3 and proceed directly to Section 3. For the other readers less familiar with complex networks, this section will provide essential material on networks required for understanding the subsequent sections.

### 2.1. Terminology and Notation

In this review, we will denote the total number of nodes in a network as $N$ and the total number of edges as $M$. Networks are generally classified as *undirected* or *directed* corresponding to those that consist of undirected or directed edge interactions, respectively. Nodes are labeled by Latin letters such as $i$ and $j$, undirected edge between $i$ and $j$ is denoted by $(i,j)$, while a directed edge from $i$ to $j$ is denoted by $\langle i,j \rangle$. *In-links* and *out-links* are used to describe directed edges going in and out of a node. Most networks do not have parallel edges, i.e., edges with the same pair of end nodes and directions. The connectivity between nodes are characterized by the *adjacency matrix* $A$, with symmetric entries $a_{ij} = a_{ji} = 1$ when an undirected edge exists between nodes $i$ and $j$, and $a_{ij} = 1$ in the case of a directed edge from node $i$ to $j$; and 0 otherwise. We denote by $\mathcal{L}_i$ the set of nodes connected to $i$. The diagonal elements $a_{ii}$ in $A$ are set to zero in most cases if link from node $i$ to itself does not exist. Networks are said to be *sparse*, or sparsely connected, when nodes are generally connected to a small number of neighbors compared to $N$, and *dense*, or densely connected when the number of connections per node is $O(N)$. In the study of disordered systems, the misleading term *extreme dilution* is sometimes used to describe a connectivity level smaller than $O(N)$ but much larger than $O(1)$.

*2.2. Features and Measures*

A number of prominent network features have been observed and identified, and common measures have been introduced to quantify them. They help reveal the significance of individual nodes and edges, and the statistical and topological properties of networks. Some of these measures such as node degree and edge betweenness are of high relevance to networking systems.

*2.2.1. Node and Edge Centrality*

Node and edge centrality correspond to the significance of individual nodes and edges in a network. The simplest yet important centrality measure for nodes is their *degree*. For undirected networks, the degree of a node is defined as the number of edges connected to it, and is usually denoted as $k_i = \sum_j a_{ij}$ for node $i$. For directed networks, edges linking a node can be either going into or out of it, and the *in-degree* and *out-degree* of a node are defined as the number of in-links $k_i^{\text{in}} = \sum_j a_{ji}$ and out-links $k_i^{\text{out}} = \sum_j a_{ij}$, respectively. In general, nodes with degrees significantly larger than the surrounding nodes are considered important in the network, and are sometimes called hubs.

Another common measure for node centrality is the node betweenness, a measure based on the fraction of shortest paths between any pair of nodes that passes through the node of interest (note that the shortest path is not always unique and in many cases more than one shortest path exist between a particular pair of nodes.) Nodes are said to have high betweenness if a substantial fraction of the shortest paths pass through them, and hence are considered to be important, in particular for networking systems [42]. Denoting the number of shortest paths from $x$ to $y$ that pass through node $i$ as $n_{xy}^i$, the normalized betweenness $B_i$ of node $i$ is given by

$$B_i = \frac{1}{(N-1)(N-2)} \sum_{\substack{x,y \\ x \neq y \neq i}} \frac{n_{xy}^i}{n_{xy}}, \tag{1}$$

where $n_{xy}$ is the total number of shortest paths from $x$ to $y$ and $(N-1)(N-2)$ is the number of unordered node pairs in an undirected network. The above expression is normalized such that when all the shortest paths between all node pairs pass through $i$, $B_i = 1$. We note that generally $n^i_{xy} \neq n^i_{yx}$ and $n_{xy} \neq n_{yx}$ in directed networks, while in undirected networks $n^i_{xy} = n^i_{yx}$ and $n_{xy} = n_{yx}$ and the computations of $B_i$ can be simplified by summing only unordered node pair $(x, y)$ and multiplying the expression by two. A similar measure on edges is known as edge betweenness, which counts the fraction of shortest paths between any node pair that passes through the edge of interest. Given $n^{\langle i,j \rangle}_{xy}$ is the number of shortest paths from $x$ to $y$ that pass through the edge $\langle i, j \rangle$, the normalized betweenness of edge $\langle i, j \rangle$ is defined as

$$B_{\langle i,j \rangle} = \frac{1}{N(N-1)} \sum_{\substack{x,y \\ x \neq y, i \neq j}} \frac{n^{\langle i,j \rangle}_{xy}}{n_{xy}}, \quad (2)$$

where $n_{xy}$ is again the total number of shortest paths from $x$ to $y$. For an undirected edge $(i, j)$ in undirected networks, $B_{(i,j)} = B_{\langle i,j \rangle} = B_{\langle j,i \rangle}$ and computations can again be simplified by summing only unordered node pairs $(x, y)$ and multiplying the expression by two.

### 2.2.2. Statistical Properties

While node and edge centralities are local measures on individual nodes and links, the statistical properties of networks measure macroscopic features and are often referred to as the network characteristics. For instance, the degree distribution $p(k)$ is an important statistical property which characterizes the probability that a randomly chosen node is of degree $k$; it facilitates measuring and effectively categorizing the network type and features. For instance, *exponential networks* correspond to networks where degree distribution is exponentially bounded, i.e., decays exponentially or faster. We will see later on, in Section 2.3, that random networks [43, 44] and model networks exhibiting small world phenomenon [45] belong to the class of exponential networks. On the other hand, networks characterized by a power-law degree distribution are often called *scale-free networks*. A substantial amount of effort has been dedicated to the mechanism by which networks grow to a given degree distribution [15]. In networking systems, which are the focus of this review, the form of degree distribution has a large influence on traffic congestion [42, 46].

Another useful macroscopic measure, in addition to the degree distribution, is the correlation between degrees of neighboring nodes. We denote by $p(k'|k)$ the conditional probability that a node of degree $k'$ is connected a neighbor of degree $k$. *Uncorrelated networks* are networks where degrees of neighbors are uncorrelated, which gives rise to a $k$-independent distribution for $k'$, $p(k'|k) = k'p(k')/\langle k \rangle$, equivalent to the probability of a randomly chosen edge to be connected to node with degree $k'$. The average degree of nearest neighbors of a node of degree $k$ can be expressed as [47]

$$k^{\text{nn}}(k) = \sum_{k'} k' p(k'|k), \quad (3)$$

providing a simple empirical measure of node correlation. When $k^{\text{nn}}(k)$ is independent of $k$, the network is termed uncorrelated; in which case $k^{\text{nn}}(k) = \langle k^2 \rangle / \langle k \rangle$ [48]. On the other hand, when $k^{\text{nn}}(k)$ depends on $k$, node degrees are correlated. An increasing $k^{\text{nn}}(k)$ indicates that nodes with large degrees are connected to each other and the network is called *assortiative*; while a decreasing $k^{\text{nn}}(k)$ indicates that nodes with higher degrees are connected to nodes with lower degrees and the network is called *dissortiative*.

An alternative measure for assortiativity is the assortiativity coefficient introduced by Newman [49]. To evaluate assortiativity, one first defines the *excess degree* of node $i$ to be $z_i = k_i - 1$, i.e. one less than the actual degree of node $i$. One then defines $q(z)$ to be the probability of arriving at a node with excess degree $z$ from a random link, given by

$$q(z) = \frac{(z+1)\,p(z+1)}{\sum_k k\,p(k)} = \frac{(z+1)\,p(z+1)}{\langle k \rangle}, \tag{4}$$

where $p(k)$ and $q(z)$ are the degree and excess degree distributions, respectively. The first factor in the product of $(z+1)p(z+1)$, i.e. $k\,p(k)$, corresponds to the fact that random walkers are more likely to arrive at nodes with large degrees as they have more edges. We remark that for Erdös-Rényi (ER) networks described in Sec. 2.3.1, $q(k) = p(k)$. The assortiativity coefficient is then given by

$$R = \frac{1}{\sigma_q^2} \sum_{z,z'} zz'[q(z,z') - q(z)q(z')] \tag{5}$$

where $q(z, z')$ is the probability of a randomly chosen edge with excess degrees $z$ and $z'$ at the end nodes. The normalization constant $\sigma_q$ is the variance of the distribution $q(z)$ as given by $\sigma_q = \sum_z z^2 q(z) - [\sum_z zq(z)]^2$. The term $q(z)q(z')$ describes the probability when the occurrences of $z$ and $z'$ on an edge are independent. The assortiativity coefficient $R$ is thus the Pearson coefficient [50] on the correlation between $z$ and $z'$; $R = 1$ when $z$ and $z'$ are completely correlated and $R = 0$ when no correlation exists between $z$ and $z'$.

Correlation measures on three nodes were also introduced, in addition to the more commonly used two-node measures. For instance, the *clustering coefficient* measures the popularity of triadic relations/interactions, i.e. the occurrence of triangles in networks. Clustering coefficients can be defined locally on individual nodes as follows.

Note that the maximum number of links that can appear in the nearest neighborhood of a node is $k(k-1)$ in directed networks and $k(k-1)/2$ in undirected networks. The clustering coefficient $C_i$ at node $i$ is defined as the ratio of the number of existing links in the neighborhood of a node to the maximum possible number

$$C_i = \frac{1}{k_i(k_i - 1)} \sum_{\substack{x,y \in \mathcal{L}_i \\ x \neq y}} a_{xy} \tag{6}$$

where $\mathcal{L}_i$ corresponds to the set of nearest neighbors of node $i$.

As for undirected networks, one can compute $C_i$ by summing only unordered node pairs $(x, y)$ and then multiply the summation by two. A unity clustering coefficient for node $i$ corresponds to the case where all nearest neighbors of $i$ are connected to each other. The network clustering coefficient $C$ is just the average of the local clustering coefficient at all individual nodes $C = \langle C_i \rangle$. Note that the clustering coefficient measures the local triadic relations at a node, and a high clustering coefficient does not reflect the global clustering of nodes into communities.

*2.2.3. Topological Features*

Other than statistical properties, the topological structure of networks may show characteristic features that are not necessarily captured by the former. One of the most important topological features is the emergence of communities, determined by an increased intra-community link density with respect to the lower inter-community density. Communities typically appear at the mesoscopic scale as they are often small compared the network size but larger than a few individual nodes. It has been shown that the Internet exhibits strong community structure corresponding to individual countries or regions [51]. This feature is also observed in social networks, where people are sometimes linked to others who share similar interests, or are in the same social group [52, 53].

It is often easier to quantify community structure in undirected networks, while the community detection in directed networks remains an area of research [54, 53]. Given a specific grouping of nodes, a common measure to quantify the strength of community structure is known as the modularity $Q$ [55], defined as

$$Q = \frac{1}{2M} \sum_{\substack{i,j \\ i \neq j}} \left( a_{ij} - \frac{k_i k_j}{2M} \right) \delta_{g_i, g_j} \qquad (7)$$

where $g_i$ corresponds to the group label of node $i$. The quantity $k_i k_j / 2M$ corresponds to the expected number of edges between $i$ and $j$ in the absence of community structure. The modularity measure thus compares the density of links within the same group to the null model where nodes are connected regardless of communities. The higher the modularity, the more prominent is the community structure.

Another important topological feature is the emergence of hierarchical structures. The most typical examples are tree networks such as taxonomy scheme of animal species and pedigree chart, where nodes are organized in layers with ancestors and descendants. Hierarchical structures in such networks are represented by the group of leaf-nodes connected through the same branch. However, most networks do not have a strict tree-like topology but instead possess an underlying skeleton or backbone, which is a minimal spanning tree formed by significant network edges. Examples include the World Wide Web, the Internet and citation networks [56, 57].

There are several suggested methods to identify and quantify the hierarchical structure of networks. For instance, one can search for hierarchical paths between any two nodes by following paths where degrees first go up and then down. If a large fraction of shortest paths between randomly chosen nodes are hierarchical, the network is said to have a hierarchical structure [58]. To find the backbone of the network, one can search for the spanning tree where total edge betweenness is maximized [59]. It was shown that in such backbone networks major statistical properties such as exponents of degree distribution and betweenness distribution are preserved. Another way to quantify hierarchy is by $k$-shell decomposition, for instance, in the autonomous system (AS) level of the Internet [60]. We remark that in some cases, networks are not completely modular or hierarchical, but instead show a mixture of characteristics.

*2.2.4. Paths*

Paths between nodes correspond to the contiguous set of edges constituting a route from one node to another. In addition to the betweenness measure, which was described earlier and involves shortest paths, other network properties are used to evaluate the distance between nodes. For instance, the *average path length* is a commonly used measure to determine the expected level of separation between network nodes. Another measure of proximity between network nodes is the *maximum length* of shortest paths among all node pairs, indicating how far apart remote nodes are; this is sometimes termed the *network diameter*. Effective search strategies for finding shortest paths are crucial in networking applications [61].

An example where path length is employed to characterize node distance is the well-known six degree of separation paradigm in social networks [62], which suggests that any two people in the world are separated, on average, by a sequence of six intermediate contacts (people are represented as nodes, acquaintances as edges). Social network is thus a small-world network, to be determined formally in Section 2.3, which corresponds to networks with average path length that scales as $\log N$, compared to regular lattice where the average shortest distance scales as $N$. We note that random regular graphs and Erdös-Renyi networks (random graphs) are small-world in this sense as their average shortest distance also scales as $\log N$ [63]. Other examples of small-world networks include the power-grid and actor collaboration networks [45].

In directed networks, there are typically pairs of nodes which are not connected by directed paths, where all individual edges follow the same direction, making path length undefined. To evaluate the average path length and diameter in directed networks, edges are sometimes regarded as undirected. A sub-graph in a directed network where all node pairs are connected by directed path is known as a *strongly connected component*. Random walk, an important tool in the exploration of networks, which will be formally introduced in Section 3, is only guaranteed to coverage on strongly connected networks.

*2.3. Network Models*

Various network models have been introduced to study and understand the origin of network features and structure. While many of these models only describe the mechanism of link formation some suggest also a mechanism for node addition aimed to explain how networks grow. Understanding these mechanisms is crucial to networking, especially the resilience to failure and attack described in Section 3.2 and path optimization described in Section 4.3. In this section, we will briefly review some of more popular network models and the corresponding implications.

*2.3.1. Erdös-Rényi Networks*

Erdös-Renyi networks are often referred to as random graphs or ER networks. The most commonly studied version, in which links appear randomly with equal probability, was first introduced by Gilbert [43]; a closely related variation had been later introduced and studied by Erdös and Rényi [44]. In an ER network with $N$ nodes, an undirected edge exists between any node pair with equal probability $p$ independently of other existing edges. The expected number of edges in ER networks is $pN(N-1)/2$, and the node degrees follow a binomial distribution which is effectively approximated by a Poisson distribution with an average $pN$ when the network size $N$ is large, given by

$$p(k) = \frac{(pN)^k}{k!}e^{-pN}. \tag{8}$$

The probability distribution $p(k)$ has a prominent peak at the expected value $\langle k \rangle$ and decays faster than exponential for $k > \langle k \rangle$, which classifies ER as exponential networks. While $p = 1/N$ is the usual percolation threshold for the network to have a largest connected component of size $O(\ln N)$, it has been shown that when $p > \ln(N)/N$ the whole network tends to be connected with no isolated nodes [63]. The properties of ER networks such as connectedness and the size of connected components, have been extensively studied by graph theorists; ER networks are often referred to as random graphs and are denoted as $G_{N,p}$ [63].

For researchers studying phenomena on complex networks, ER networks are a useful modeling tool and are considered as a benchmark for comparison with other network models. Since edges are assigned randomly in ER networks, they do not show any community or hierarchical structures.

*2.3.2. Scale-free Networks*

Scale-free networks correspond to networks with a power-law degree distribution, of the form $p(k) = ck^{-\gamma}$, where $c$ is a normalization constant. Many of the networks one encounters in daily life are characterized by a power-law degree distribution, and thus are scale-free. Scale-free networks are characterized by a high number of hubs (i.e., highly connected nodes) which results from the long tail in the degree distribution, compared to the Poissonian distribution in ER networks.

The most studied model which generates a power-law degree distribution is the Barabási-Albert (BA) model [64]. The BA model starts initially with $m_0$ nodes, with new nodes introduced at each time step. The latter are connected to $m$ existing nodes with a probability proportional to their degrees, i.e. nodes of a large degree are more likely to be connected. This growth method is often known as preferential attachment or the rich-get-richer effect; the concept has been used earlier to explain power-laws by Yule [65] and Simon [66], and is sometimes called the Yule's process. With regard to networks, Barabási and Albert showed that growth and preferential attachment are the two essential ingredients leading to networks with a power-law distribution, and thus the BA model is often employed for studying phenomenon based on network growth. To obtain the degree distribution analytically, one can use the recursive equation [15, 67]

$$p(k, s, t+1) = \frac{k-1}{2t} p(k-1, s, t) + \left(1 - \frac{k}{2t}\right) p(k, s, t), \qquad (9)$$

which describes the probability $p(k, s, t)$ of a node introduced at time $s$ to have a degree $k$ at time $t$. The factor $k/2t$ corresponds to the probability that a node of degree $k$ is chosen by the new node to establish a link at time $t$. By taking the stationary state limit $p(k) = \lim_{t \to \infty} \frac{1}{t} \sum_s p(k, s, t)$, Eq. (9) leads to derive a recursion relation relating $p(k)$ and $p(k-1)$, resulting in the following expression for $p(k)$

$$p(k) = \frac{2m(m+1)}{k(k+1)(k+2)}. \qquad (10)$$

For large $k$, $p(k) \propto k^{-3}$ independent of $m$ and $m_0$, the number of existing nodes at each time and initial number of nodes, respectively.

Though the BA networks are characterized by a power-law degree distribution, in most real networks the decay exponents are different from 3. Hence, mechanisms which produce a tunable exponent have been studied. For instance, the original Yule's process is employed in [67]; it is similar to the BA model except that the probability of a new node connecting to an existing node of degree $k$ is proportional to

$$\frac{k+a}{(2m+a)t}, \qquad (11)$$

where $a$ is a model parameter. It has been shown that the degree distribution at steady state is given by [67]

$$p(k) \propto k^{-(3+\frac{a}{m})} \qquad (12)$$

with an adjustable exponent. The BA model corresponds to the case of $a = 0$. Nevertheless, the exponent value of $\gamma = -3$ does mark the critical point below which the second moment $\langle k^2 \rangle$ diverges due to the divergence of $\sum_k k^{-n}$ when $n \leq 1$. We will see in section 3.2 that the divergence of $\langle k^2 \rangle$ has many important consequences in epidemic spreading and percolation, which make networks with $\gamma \leq 3$ substantially different from those with $\gamma > 3$.

Another simple approach to generate networks with power law degree distributions is the configuration model (CM) [48] which assigns a degree for every node from a predefined degree distribution $p(k) \propto k^{-\gamma}$. With the constraint of $\sum_i k_i$ being even, edges are added randomly to fill up all the histogram of $k_i$ values. In general, CM generates networks with any prescribed degree distribution. A different approach to generate networks with power law distribution is the good-get-richer mechanism [68] in which nodes are characterized by an intrinsic fitness, and the existence of links between a pair of nodes is proportional to the product of their fitnesses. Galderalli et al [68] showed that power-law degree distributions with varying exponents are generated by a power-law or exponential distribution of fitness. Combinations of the preferential attachment and good-get-richer effect have been also studied to obtain varying exponents of the power-law degree distributions [69].

*2.3.3. Small World Networks*

The small world network model was introduced by Watts and Strogatz to model the small world phenomenon observed in social networks and is often referred to as WS network [45]. To model networks where average path length scales as $\log N$, Watts and Strogatz start from a regular ring of $N$ nodes, each with $k$ neighbors in its nearest neighborhood. Each edge is then rewired with probability $p$ to a random node. The case with $p = 0$ corresponds to the original regular ring while the case with $p = 1$ corresponds effectively to a random network. For intermediate values of $p$, the networks show small average path length $L$ but large clustering coefficient $C$, compared to the large coefficients $L$ and $C$ in regular network and small ones in random networks. These results show that a small value of $p$, i.e. a small number of long-range edges would substantially decrease average path length while keeping local clustering. Long-range edges which connect nodes that were originally far apart are usually of high betweenness and are essential to the small world phenomenon. WS networks have a degree distribution $p(k) = \delta_{k,\langle k \rangle}$ when $p = 0$ and a distribution $p(k)$ similar to ER networks when $p = 1$; WS networks are thus considered to be within the family of exponential networks. It has been shown that BA networks also exhibit small world properties, as their average path length scales approximately as $\log N$ [15].

## 3. Networking as a Dynamical Process

After introducing measures and models to characterize network structure, we will review network dynamical processes. They play an important role in tackling various networking-related problems and searching for optimal networking solutions, for instance, routing and searching protocols and the containment of computer viruses. In this section, we briefly describe processes that include random walk, epidemic spreading and cascading failures on networks. We remark that epidemic spreading and cascading failures have been included in the comprehensive review by Boccaletti et al [19]. While we will briefly describe essential concepts and processes, we will focus on more recent developments in the two areas which are not included in [19], for instance cascading failures on interdependent networks.

*3.1. Random Walk*

Random walk corresponds to a sequence of random movements in successive time steps. It was first studied in one dimension to describe a random walker hopping left and right randomly on a straight line. Given an equal probability for hopping left or right and an equal distance $d$ on each hop, the position of the random walker from the starting point after $t$ hops follows a Gaussian distribution with zero mean and variance $td^2$, i.e. the expected distance of the random walker from the starting point is proportional to $\sqrt{t}d$. It describes a Markov process as, given the present move of the random walker, the probability of his next position is independent of the history. The analysis of random walks on regular lattices was later extended to higher dimensions such as 2D and 3D square lattices as well as on general networks to model different physical phenomena, for instance, Brownian motion. Some of these studies such as the scaling properties [70] and their relations to the spectra of networks [71] are of high relevance to networking applications. The same methodology has been adopted by other disciplines; in economics to model the dynamics of stock price; in biology to describe the movement of molecules in cells; in ecology to model the predator-prey ecosystem and in computer science for state-space sampling.

In the context of networking, random walk is useful for modeling physical and cyber transport processes, especially the delivery of information from one node to another. Knowledge of its dynamics, such as the first passage time, return time and cover time [72, 73, 74] are valuable for developing local search strategies. Local search is typically computationally efficient as it utilizes only local information as compared to centralized search algorithms which utilize global information. Consequently, the computational complexity of random walks scales favorably with the system size, which makes it a potentially efficient tool for searching large networks such as the World Wide Web. Thus, random walk has been modified and applied to rank the significance of webpages in the World Wide Web [75], to route packets from sources to destinations [76], to identify communities in directed networks [77], to devise local search algorithms in file sharing systems [78] and to mitigate congestions in communication networks [79].

### 3.1.1. Equilibrium Properties and PageRank

A random walk can be represented by a master equation which describes the probability of a random walker to be at a certain network node. We first denote $\pi_i(t)$ as the probability that a random walker would arrive at node $i$ after $t$ time steps. The initial condition of the walker is given by $\pi_i(0)$, for instance, $\pi_i(0) = 1/N$ when the initial position of the walker is evenly distributed on all network nodes. The master equation which describes the dynamics of $\pi_i(t)$ is then given by

$$\pi_i(t+1) = \sum_{j=1}^{N} \frac{a_{ji}}{k_j} \pi_j(t) \tag{13}$$

for undirected networks. The factor $a_{ji}/k_j$ represents the transition probability from state $j$ to $i$, where $a_{ji}$ and $k_j$ are the adjacency matrix element and the degree of node $i$, respectively. The same equation applies to directed networks by replacing $k_j$ in Eq. (13) by $k_j^{\text{out}}$.

Equivalently, Eq. (13) can be represented by matrix multiplication. We first denote $\vec{\pi}(t)$ to be a vector with elements $\pi_i(t)$ for $i = 1 \cdots N$. Different initial conditions are represented by the choice of $\vec{\pi}(0)$. The dynamics of $\vec{\pi}(t)$ is then given by

$$\vec{\pi}(t+1) = \mathcal{M} \cdot \vec{\pi}(t) \tag{14}$$

where $\mathcal{M}$ is a left stochastic transition matrix (i.e. column-normalized) with elements $m_{ji} = a_{ji}/k_j$ for undirected networks and $m_{ji} = a_{ji}/k_i^{\text{out}}$ for directed ones.

Other than merely a representation, Eq. (14) allows one to determine the convergence condition for $\vec{\pi}(t)$ when $t \to \infty$. By Perron-Frobenius theorem, if $\mathcal{M}$ is irreducible and primitive, then as $t \to \infty$ the state probability $\vec{\pi}(t)$ converges to the unique eigenvector of $\mathcal{M}$ with eigenvalue 1. The irreducibility and primitivity of the matrix $\mathcal{M}$ has a strong correspondence to its underlying network structure; $\mathcal{M}$ is irreducible if the network is strongly connected, i.e., a path exists from any node to any node. This condition is readily satisfied for a connected undirected network, but for directed networks there are typically node pairs with no directed path from one node to another. To satisfy convergence requirements of a random walk in directed networks, transitions between all nodes should be allowed, which is essential for search algorithms such as *PageRank* [75].

The primitivity of a matrix $\mathcal{M}$ corresponds to all elements of $\mathcal{M}^n$ being positive for some natural number $n$. In terms of topology, a network is primitive if all node pairs can be linked by exactly $n$ hops. To demonstrate that primitivity is not always satisfied we provide a simple example of an irreducible but non-primitive network with three nodes and three directed edges, forming a directed triangle. The network is not primitive as node pairs are connected by either odd or even number of hops, such that $\mathcal{M}^n$ always has zero elements for any $n$, and the random walk probability $\vec{\pi}(t)$ does not converge for some initial conditions.

While it is difficult to show in general that the convergence condition is met, a steady state probability $\vec{\pi}(t)$ can be easily obtained at least for undirected networks. It is straightforward to show that the vector $\vec{\pi}(\infty)$ with elements

$$\pi_i(\infty) = \frac{k_i}{\sum_j k_j} = \frac{k_i}{N\langle k \rangle} \tag{15}$$

is an eigenvector of $\mathcal{M}$ in Eq. (14) with eigenvalue 1, where $\sum_j k_j$ is a normalization constant. This can be readily justified by a direct substitution of $\pi_j(\infty)$ into Eq. (13) when $t \to \infty$, which yields

$$\pi_i(\infty) = \sum_{j=1}^{N} \frac{a_{ji}}{k_j} \pi_j(\infty) = \sum_{j=1}^{N} \frac{a_{ji}}{k_j} \frac{k_j}{N\langle k \rangle} = \frac{k_i}{N\langle k \rangle} \ . \tag{16}$$

In other words, Eq. (15) implies that in an undirected network the probability that a node hosts the random walker in the steady state is proportional to its degree. The same is not necessarily true for directed networks, as the sum in Eq. (13) implies that $\pi_j(\infty)$ is roughly proportional to $k_j^{\text{in}}$, which generally does not cancel the denominator of $k_j^{\text{out}}$ in Eq. (16) for directed networks. Consequentially, the steady state of the random walker probability is less obvious in directed networks.

Here we briefly describe the *PageRank* algorithm [75] which utilizes the steady-state random walker probability in directed networks to rank webpages. PageRank forms the basis of the Google search engine and models a surfer clicking randomly through the hyperlinks on the World Wide Web. In general, the higher the probability of a random walker arriving at a web page, the higher is the rank given to the page. The underlying mechanism is identical to the traditional random walk except that the web surfer occasionally moves randomly to another website regardless of connectivity; Eq. (14) is then modified to become

$$\vec{\pi}(t+1) = \frac{\lambda}{N} + (1-\lambda)\mathcal{M}' \cdot \vec{\pi}(t) \tag{17}$$

where the elements of $\mathcal{M}'$ are given by

$$m'_{ji} = (1 - \delta_{k_j^{\text{out}},0})\frac{a_{ji}}{k_j^{\text{out}}} + \delta_{k_j^{\text{out}},0}\frac{1}{N} , \qquad (18)$$

which is identical to $\mathcal{M}$ except for columns that represent nodes with zero out-degree. The *return probability* $\lambda$ represents the probability of the surfer to return to a random page. As mentioned before, a non-zero $\lambda$ would introduce possible transitions among nodes which make the network effectively strongly connected, and thus the PageRank algorithm is likely to converge. Efforts to analytically approximate PageRank's asymptotic probability generally show that $\pi_i(\infty) \propto k_i^{\text{in}}$ [80].

### 3.1.2. First Passage Time and Coverage Dynamics

Apart from the steady state and its use in PageRank, the dynamics of a random walk is of great importance for networking applications. The most studied dynamical quantities include the mean first return time (MFRT) and the mean first passage time (MFPT). We denote by $\langle T_{i \to j} \rangle$, the average time required for a random walker who started at node $i$ to first arrive at node $j$ on a given graph. The MFRT $\langle T_{i \to i} \rangle$ is the average time for the random walker to return to $i$ after a random walk, while the MFPT $\langle T_{i \to j} \rangle$ is the mean time required to get from $i$ to $j$. We note that in general $\langle T_{i \to j} \rangle \neq \langle T_{j \to i} \rangle$. Following Kac's formula [81, 82], which describes the recurrent probability of discrete stochastic processes, MFRT of a random walk is given by

$$\langle T_{i \to i} \rangle = \frac{1}{\pi_i(\infty)}, \qquad (19)$$

where $\pi_i(\infty)$ is the steady state probability of a random walker to be at node $i$. An alternative proof of Eq. (19) will be given below as part of the derivation of MFPT.

The MFPT is clearly dependent on the distance between nodes $i$ and $j$ and on the network topology. Noh and Rieger derived a general expression for MPFT which is applicable for arbitrary networks [72]. Here we briefly describe their derivation. To obtain the expression for MPFT, one starts with the relation

$$\pi_{i \to j}(t) = \delta_{t,0}\delta_{i,j} + \sum_{t'=0}^{t} \pi_{j \to j}(t-t')f_{i \to j}(t') \qquad (20)$$

where the first term enforces the initial condition $\pi_{i \to j}(0) = \delta_{i,j}$; and $f_{i \to j}(t')$ denotes the probability that a random walker starting at node $i$ first arrives at node $j$ after $t'$ time steps. We note that the MFPT from $i$ to $j$ is given by [72]

$$\langle T_{i \to j} \rangle = \sum_{t=0}^{\infty} t f_{i \to j}(t) = -\tilde{f}'_{i \to j}(0), \qquad (21)$$

with $\tilde{f}_{i \to j}(s)$ being the Laplace transform $\tilde{f}(s) \equiv \sum_{t=0}^{\infty} e^{-st}f(t)$. To evaluate the derivative $\tilde{f}'_{i \to j}(0)$, one considers the Laplace transform of Eq. (20) and expresses $\tilde{f}_{i \to j}(s)$ as

$$\tilde{f}_{i \to j}(s) = \frac{\tilde{\pi}_{i \to j}(s) - \delta_{i,j}}{\tilde{\pi}_{j \to j}(s)}. \qquad (22)$$

Finally, one defines the relaxation moment of $\pi_{j \to j}$ as $r_{ij}^{(n)} = \sum_{t=0}^{\infty} t^n [\pi_{i \to j}(t) - \pi_j(\infty)]$ and expresses $\tilde{\pi}_{i \to j}(s)$ as

$$\tilde{\pi}_{i \to j}(s) = \frac{\pi_j(\infty)}{1 - e^{-s}} + \sum_{n=0}^{\infty} (-1)^n r_{ij}^{(n)} \frac{s^n}{n!}, \quad (23)$$

which can be readily justified by substituting the definition of $r_{ij}^{(n)}$. The combination of Eqs. (22) and (23) would help us to evaluate $\tilde{f}'_{i \to j}(0)$, and thus the MFPT

$$\langle T \rangle_{i \to j} = \begin{cases} \dfrac{N \langle k \rangle}{k_j}, & \text{for } j = i \\ \dfrac{N \langle k \rangle}{k_j} [r_{jj}^{(0)} - r_{ij}^{(0)}]. & \text{for } j \neq i \end{cases} \quad (24)$$

Since $\pi_i(\infty) = k_i / N \langle k \rangle$ for undirected network as given by Eq. (15), $\langle T_{i \to i} \rangle = N \langle k \rangle / k_i = 1/\pi_i(\infty)$, which is the expression for first return time of Eq. (19). We refer the reader to [72] for the detailed derivation.

As $r_{jj}^{(0)}$ and $r_{ij}^{(0)}$ are governed by the random walk relaxation in the whole network, there are no explicit expressions for Eq. (24). However, Noh and Reiger were able to draw a physical interpretation and insight from the difference $\langle T_{ij} \rangle - \langle T_{ji} \rangle$ [72]. By using Eq. (24), $\langle T_{ij} \rangle - \langle T_{ji} \rangle = c_j^{-1} - c_i^{-1}$ with

$$c_i = \frac{k_i}{r_{ii}^{(0)} N \langle k \rangle}, \quad (25)$$

which they call the *random walk centrality*. It implies that when a node $i$ has a large random walk centrality, the difference $\langle T_{ij} \rangle - \langle T_{ji} \rangle$ tends to be more negative, i.e. the mean first passage time is shorter from the rest of the network to $i$ than from $i$ to the the rest of the network. They show that in scale-free network, the variation of $r_{ii}^{(0)}$ across nodes is small compared to the degree $k$. Hence, the random walk centrality is mainly governed by $k$, and nodes with a higher degree would be more efficient in receiving information as they have a larger random walk centrality. This is important for search algorithms, which will be discussed in the next subsection.

A simple way to further understand Eq. (24) was suggested in [83]. Consider a random walker walking on a uncorrelated network from $i$ to $j$, the MFPT can be estimated by

$$\langle T \rangle_{i \to j} \approx \sum_{t=1}^{\infty} t [1 - \pi_j(\infty)]^{t-1} \pi_j(\infty) \quad (26)$$

where $[1 - \pi_j(\infty)]^{t-1} \pi_j(\infty)$ is the probability of a random walk through a path of nodes other than $j$ for $t-1$ steps and arriving at $j$ at time $t$. With $\pi_j(\infty) = k_j / N \langle k \rangle$ in undirected networks (see Eq. (15)), the expression in Eq. (26) can be simplified to

$$\langle T \rangle_{i \to j} \approx \frac{N \langle k \rangle}{k_j}, \quad (27)$$

which reproduces the $i$-independent components in Eq. (24). Physically, this approximation is independent of the starting node $i$ and is inversely proportional to $\pi_j(\infty)$, implying that the higher the probability of the random walker to be found at $j$ in the steady state, the shorter the MFPT to $j$ is.

Another interesting dynamical quantity is the coverage $S(t)$, defined as the fraction of nodes visited by the random walker by time $t$. Baronchelli et al [83] provide a simple argument to derive an approximate form of $S(t)$. One first denotes the fraction of nodes of degree $k$ visited by the random walker by time $t$ as $s_k(t)$; the rate of change of $s_k(t)$ can be written as

$$\frac{\partial s_k(t)}{\partial t} = [1 - s_k(t)]k \left[ \sum_{k'} p(k'|k) \frac{\rho_{k'}(t)}{k'} \right], \tag{28}$$

where $\rho_k(t)$ is the probability that a random walker arrives at a vertex of degree $k$ at time $t$. The first term on the right hand side is the fraction of newly-visited nodes with degree $k$ and the second term is the random walker probability of leaving nodes of degree $k$. By approximating $\rho_{k'}(t) \approx k'/N\langle k \rangle$, using the steady-state values, one obtains

$$\frac{\partial s_k(t)}{\partial t} = [1 - s_k(t)] \frac{k}{N\langle k \rangle}, \tag{29}$$

such that

$$s_k(t) = 1 - \exp\left(-\frac{kt}{N\langle k \rangle}\right). \tag{30}$$

The coverage $S(t)$ is then given by $S(t) = \sum_k p(k) s_k(t)$. We can see that as $t \to \infty$, $S(t) \to 1$, implying a full coverage of the network by the random walker. At intermediate values of $t$, $kt/\langle k \rangle N \ll 1$ and the expansion of Eq. (30) leads to $S(t) \propto t$, which is also observed for random walks on small world networks [70].

Other dynamical properties of random walks that have been studied include the relationship between passage time and distance [84], universal dynamic scaling functions in small-world networks [70] and explicit expressions for cover and communication times in specific graphs [74].

### 3.1.3. Routing and Searching by Random Walk

Random walk can be applied to model various phenomenon in networking systems. For instance, in packet-switched networks, the queuing of packets in a single buffer can be modeled by a one-dimensional biased random walk with hopping rate that depends on the arrival rate of packets [85]. On the other hand, random walk can be employed within searching and routing protocols [78, 61, 79, 86, 87] due to its efficiency in searching complex networks. Here we briefly describe the network traffic model [76] and illustrate how random walk is applied to networking.

More specifically, one defines the probability of individual nodes to generate a packet at any time step as $\lambda$. Each packet is randomly assigned a destination different from the generating node. Following a specific routing protocol, the packet then travels from its source to its destination and vanishes upon arrival; each node in the network serves as a packet generator as well as a router. At each time step, node $i$ forwards to its neighbors at most $c_i$ packets waiting in its queue; $c_i$ is thus the capacity of router $i$. In the simplest case, one can consider infinite queuing capacity and packets are forwarded in a first-in-first-out (FIFO) basis [42].

By denoting $n(t)$ as the number of undelivered packets traveling in the network at time $t$, one expects that $n(\infty)$ tends to stabilize when the packet generating rate $\lambda$ is small, corresponding to a *free-flow* state where the numbers of created and delivered packets are balanced [42]. In contrast, when $\lambda$ is large, the packets generated outnumber the packets delivered per time step and $n(t)$ increases with time. This corresponds to a *congested* phase where traffic congestions occur. To better illustrate the phenomenon, Arenas et al [76] introduced an order parameter $\eta$ which corresponds to the rate of change in the number of undelivered packets, given by

$$\eta(\lambda) = \frac{1}{N} \lim_{t \to \infty} \frac{\langle \Delta n \rangle}{\lambda \Delta t}, \tag{31}$$

where the total number of generated packets per time step is $\Delta n = n(t+\Delta t) - n(t)$ and the average $\langle \cdots \rangle$ is taken over a time interval $\Delta t$. When $\eta = 0$, the system is in the free-flow state. When $\eta > 0$, there is a net increase in the number of undelivered packets and the network is in a congested phase. With a shortest path routing protocol and a different capacity per node $c_i$, a phase change at $\lambda = \lambda_c$ (characterized by an abrupt increase in $\eta$) is observed in hierarchical networks, regular networks, ER and scale-free networks [76, 42].

To apply random walk to networking, one adopts a heuristic approach and defines $p_{j \to i}$ to be the probability that $j$ forwards a packet to $i$. When $p_{j \to i}$ takes the form of

$$p_{j \to i} = \frac{k_i^\alpha}{\sum_{i'} a_{ji'} k_{i'}^\alpha}, \tag{32}$$

where $a_{ji'} = 1$ denotes the existence of an edge between nodes $j$ and $i'$ and is zero otherwise, the packets follow a path governed by *preferential random walk*. When $\alpha > 0$, the packets are more likely to go through nodes with a high degree. An extreme case is $\alpha \to \infty$ such that packets are always forwarded to neighbors with the highest degree. It was shown that such routing protocol, though heuristic, is effective in packet transmission through paths of length proportional to its shortest path in scale-free networks [61]. Such protocol is also observed to be effective for searching in scale-free networks [78]. Despite a short routing path, we will see that traffic congestions tend to occur under such routing protocol.

To understand the physical origin of congestion one first denotes $n_i(t)$ as the number of packets waiting at $i$ at time $t$ [86] and expresses the rate of change of $n_i(t)$ in the free-flow state as

$$\frac{dn_i(t)}{dt} = -n_i(t) + \sum_{j=1}^{N} a_{ij} n_j(t) p_{j \to i}, \tag{33}$$

where the first term corresponds to node $i$ forwarding all its packets to its neighbors in the free-flow state. We note that in the case of uncorrelated networks, the denominator of Eq. (32) is roughly proportional to $k_j$, such that $p_{j \to i} \propto k_i^\alpha / k_j$. Together with the steady state condition of $dn_i(t)/dt = 0$, $n_i(\infty)$ can be written as

$$n_i(\infty) \propto k_i^\alpha \sum_j a_{ij} \frac{n_j(\infty)}{k_j}. \tag{34}$$

One applies the ansatz $n_j(\infty) \propto k_j^\theta$, with some exponent $\theta$, to find that

$$n_i(\infty) \propto k_i^\alpha \sum_j a_{ij} k_j^{\theta-1} \propto k_i^{\alpha+1}, \tag{35}$$

where we have used again the assumption of uncorrelated networks in the derivation of the last expression. The ansatz of $n_j(\infty) \propto k_j^\theta$ is self-consistent when $\theta = \alpha + 1$, implying that the number of packets in $i$ at the steady state is proportional to $k_i^{\alpha+1}$. When $\alpha$ is large, the hubs in scale-free networks may get overloaded as the number of packets increases as $k_i^{\alpha+1}$. Wang et al [86] argued that $\alpha = -1$ is an alternative routing protocol which balances the traffic load, although with a longer transmission time and path length for each packet. They show that in cases of $\alpha = -1$ and uniform capacity $c_i$, congestions are less likely to occur and the free-flow state exists for a larger packet generating rate $\lambda$.

## 3.2. Epidemic Spreading

Epidemic spreading corresponds to the spread of disease among individuals, its relevance to networking has been recently recognized by the increasing prevalence of computer viruses. For instance, Pastor-Satorras and Vespignani [88] relate the absence of epidemic threshold, defined as a critical epidemic infection/spreading rate below which an epidemic does not spread, in scale-free networks to the persistent but limited prevalence of computer viruses on the Internet, despite the presence of anti-virus softwares. On the other hand, Cohen et al [89] showed that the Internet is resilient to a crash of a large fraction of its nodes, since a spanning cluster is likely to survive. Such arguments are equivalent to the absence of percolation threshold, which is strongly related to the prevalence of epidemics in scale-free networks [90]. Thus, understanding epidemic spreading would help to identify potential obstacles to effective networking.

Two major models are proposed for studying epidemic spreading [91]. The first is called the *susceptible-infected-susceptible* (SIS) model, where healthy individuals may become infected and susceptible again. In other words, individuals are either currently susceptible or infected in the SIS model. Computer viruses can be modeled by the SIS model as infected computers may become susceptible to the same virus again. The second epidemic model is called the *susceptible-infected-removed* (SIR) model, where susceptible individuals may become infected and are then removed, which means either recovery or death. In this case, individuals are characterized by the three states, susceptible, infected and removed. The SIR model is more relevant in the public health context such as the spread of influenza, where infected individuals may acquire immunity against the disease, but is also relevant to cases where nodes become immune against computer viruses.

Both SIS and SIR models had been first studied in a completely mixed population analogous to a fully connected network, and have been later studied on random, small world and scale-free networks [88, 92, 93]. At every time step, each susceptible individual becomes infected with probability $\nu$ if they are in contact with at least one infected individual. With a probability $\delta$, each infected individual becomes susceptible in the SIS model or removed in the SIR model. In cases with effective anti-virus software, $\delta$ can be set to 1 as viruses are isolated once computers are infected [88]. We define the *spreading rate* as $\lambda = \nu/\delta$ and expect a critical spreading rate $\lambda_c$ below which the epidemic dies out quickly, and above which the diseases spread widely and become persistent. We further denote by $\rho(t)$ the fraction of infected individuals at time $t$, such that a steady-state fraction $\rho(\infty)$ would be observed in SIS networks, with $\rho(\infty) > 0$ corresponding to the case when individuals are occasionally infected. On the other hand, nodes are either ultimately healthy or removed (i.e. dead or recovered) in the SIR model, with no nodes being infected in the steady state. SIR models are in fact equivalent to bond percolation model and the widely spreading phase is equivalent to the existence of giant components, which is relevant to the resilience of the Internet against router crashes [89].

*3.2.1. Steady State of Epidemic Prevalence*

To obtain the steady state of the infected fraction, one can write down the rate of change of $\rho(t)$ as a function of $t$. For networks with $p(k)$ prominently peaked at $k \approx \langle k \rangle$, such as ER and small world networks, one can adopt a mean-field approximation [92] and express the rate of change of $\rho(t)$ in the SIS model as

$$\frac{\partial \rho(t)}{\partial t} = \nu[1 - \rho(t)]\{1 - [1 - \rho(t)]^{\langle k \rangle}\} - \delta \rho(t) \tag{36}$$

where the factor $\{1 - [1 - \rho(t)]^{\langle k \rangle}\}$ corresponds to the probability that at least one of the $\langle k \rangle$ neighbors is infected. The first term on the right hand side corresponds to the fraction of healthy individuals who become infected at time $t$, while the second term is the fraction of infected nodes who recover at time $t$. As we are interested at the critical ratio $\lambda = \nu/\delta$ above which $\rho(\infty)$ is greater than zero, we expand $\{1 - [1 - \rho(t)]^{\langle k \rangle}\}$ to the lowest order in $\rho(t)$, which simplifies Eq. (36) to [92]

$$\frac{\partial \rho(t)}{\partial t} = \nu[1 - \rho(t)]\rho(t)\langle k \rangle - \delta \rho(t). \tag{37}$$

At the steady state, $\partial \rho(t)/\partial t = 0$ which implies

$$\rho(\infty) = 1 - \frac{\delta}{\nu \langle k \rangle} = 1 - \frac{1}{\lambda \langle k \rangle}, \tag{38}$$

such that $\rho(\infty) > 0$ only if

$$\lambda \geq \lambda_c = \frac{1}{\langle k \rangle}. \tag{39}$$

For recovery rate $\delta = 1$ (i.e. infected individuals that will recover in the next step), the critical infection rate becomes $\nu_c = \lambda_c = 1/\langle k \rangle$, which coincides with the bond percolation threshold in ER networks and implies that viruses become wide-spread if more than one neighbor is infected on average.

However, for networks with highly skewed degree distribution $p(k)$, the above approximation of $k \approx \langle k \rangle$ is not valid. Instead of representing the global state by a single $\rho(t)$, one introduces the variables $\rho_k(t)$ which correspond to the fraction of infected nodes at time $t$ among individuals of degree $k$. One then employs an improved mean-field approximation [88] and rewrites Eq. (37) as

$$\frac{\partial \rho_k(t)}{\partial t} = \nu[1 - \rho_k(t)]\theta_k(t) - \delta \rho_k(t), \qquad (40)$$

where $\theta_k(t)$ corresponds to the probability that a node with degree $k$ is connected to at least one infected neighbor at time $t$. If the network is uncorrelated, $\theta_k(t)$ can be expressed as

$$\theta_k(t) = 1 - \left[1 - \frac{1}{\langle k \rangle} \sum_{k'} k' p(k') \rho_{k'}(t)\right]^k \approx \frac{k}{\langle k \rangle} \sum_{k'} k' p(k') \rho_{k'}(t), \qquad (41)$$

where $k'p(k')/\langle k \rangle$ corresponds to the distribution of arriving at a node of degree $k$ from a randomly chosen edge. We have again expanded the middle term in Eq. (41) to the lowest order in $\rho_{k'}$ as we are interested in the emergence of small positive fraction of infected nodes $\rho_k(t)$. Substitution of Eq. (41) into Eq. (40) and setting it to zero leads to the critical spreading rate

$$\lambda_c^{\text{SIS}} = \frac{\langle k \rangle}{\langle k^2 \rangle}. \qquad (42)$$

We remark that the above expression is only valid for uncorrelated network as assumed in the derivation of Eq. (41). Expression (42) reduces to Eq. (39) when $\langle k^2 \rangle = \langle k \rangle^2$, i.e. node degrees are homogeneous. A similar expression of $\lambda_c$ for the SIR model is given by

$$\lambda_c^{\text{SIR}} = \frac{\langle k \rangle}{\langle k^2 \rangle - \langle k \rangle}, \qquad (43)$$

which is obtained by bond percolation approaches [89, 90].

As mentioned in Section 2.3.2, $\langle k^2 \rangle$ diverges for scale-free networks with $\gamma \leq 3$, Eq. (42) implies that $\lambda_c = 0$ for these scale-free networks with an undesirable implication: the spread of computer viruses is persistent regardless of spreading rate as the Internet is a scale-free network with $\gamma \leq 3$ [47]. Some studies argued that this result was obtained under the assumption of uncorrelated networks while in correlated scale-free networks with no connection between hubs, a positive $\lambda_c$ has been obtained [94]. Several approaches have been suggested to deal with correlated networks, including a modified expression for $\theta_k(t)$ in Eq. (41) which depends on the degree correlation given by $p(k'|k)$ [95]. Nevertheless, it was later shown that regardless of the power-law exponent, diseases would eventually die out on *finite* scale-free networks, while in *infinite* networks diseases would become persistent [96]. We will briefly describe the theory underlying these arguments in the following subsection.

*3.2.2. Propagation Dynamics of Epidemic States*

Here we describe a simple approach to obtain a general expression for epidemic threshold on any network topology [97]. We first introduce a column vector $|\rho(t)\rangle$ (the bra-ket notation is employed here for convenience in later derivations) with elements $(\rho_1(t), \cdots, \rho_N(t))$ corresponding to the probability that individual nodes are infected at time $t$. The dynamics of the system is then given by the propagation of $|\rho(t)\rangle$ as

$$|\rho(t+1)\rangle = \mathcal{W}|\rho(t)\rangle. \tag{44}$$

where $\mathcal{W}$ is the transition matrix.

To obtain an explicit form of $\mathcal{W}$, one first denotes $\eta_i(t)$ as the probability that none of the neighbors of $i$ spreads viruses to $i$ at time $t$, $\eta_i(t)$ can be expressed as

$$\eta_i(t) = \prod_{j=1}^{N}\{(1-\nu)\rho_j(t) + [1-\rho_j(t)]\}^{a_{ij}} = \prod_{j=1}^{N}\{1-\nu\rho_j(t)\}^{a_{ij}}, \tag{45}$$

where $a_{ij}$'s are elements of the adjacency matrix, taking a value 1 or 0 when edges exist or not, respectively. We then express $\rho_i(t+1)$ as a function of $\rho_i(t)$ and $\eta_i(t)$ by

$$1 - \rho_i(t+1) = \eta_i(t)[1 - \rho_i(t) + \delta\rho_i(t)] + \frac{\delta}{2}[1-\eta_i(t)]\rho_i(t), \tag{46}$$

where the first term on the right hand side corresponds to cases when neighbors of $i$ do not spread viruses to $i$, and $i$ is either healthy or recovers at time $t$. The second term corresponds to cases where neighbors spread viruses but at the same time node $i$ recovers immediately with probability $1/2$. By substitution of Eq. (45) into Eq. (46) and expansion to the first order of $\rho(t)$, assuming $\rho(t) \ll 1$; $\rho_i(t+1)$ can be written as

$$\rho_i(t+1) = (1-\delta)\rho_i(t) + \nu\sum_{j=1}^{N} a_{ij}\rho_j(t). \tag{47}$$

In terms of $|\rho(t+1)\rangle$ and $|\rho(t)\rangle$, Eq. (47) can be rewritten as

$$|\rho(t+1)\rangle = [(1-\delta)\mathcal{I} + \nu\mathcal{A}]|\rho(t)\rangle, \tag{48}$$

where $\mathcal{I}$ and $\mathcal{A}$ are the identity and adjacency matrices, respectively. Equation (48) implies that the transition matrix in Eq. (44) is $\mathcal{W} = (1-\delta)\mathcal{I} + \nu\mathcal{A}$.

To make use of the transition matrix, we first note that eigenvectors of $\mathcal{A}$ are eigenvectors of $\mathcal{W}$. By denoting $\Lambda_{\mathcal{A},n}$ and $|u_{\mathcal{A},n}\rangle$ as the $n$-th largest eigenvalue of $\mathcal{A}$ and its corresponding right eigenvector, respectively, we have

$$\mathcal{W}|u_{\mathcal{A},n}\rangle = [(1-\delta)\mathcal{I} + \nu\mathcal{A}]|u_{\mathcal{A},n}\rangle = (1-\delta-\nu\Lambda_{\mathcal{A},n})|u_{\mathcal{A},n}\rangle, \tag{49}$$

from which we see that the $n$-th largest eigenvalue of $\mathcal{W}$ is given by

$$\Lambda_{\mathcal{W},n} = 1 - \delta - \nu\Lambda_{\mathcal{A},n}. \tag{50}$$

We can now express $\mathcal{W}$ raised to power $t$ in terms of its eigenvalues and its left and right eigenvectors as

$$\mathcal{W}^t = \sum_{n=1}^{N} \Lambda_{\mathcal{W},n}^t |u_{\mathcal{A},n}\rangle\langle u_{\mathcal{A},n}|, \tag{51}$$

which is the propagator matrix from the initial state $|\rho(0)\rangle$ to the state $|\rho(t)\rangle$ at time $t$, i.e.

$$|\rho(t)\rangle = \mathcal{W}^t |\rho(0)\rangle = \sum_{n=1}^{N} \Lambda_{\mathcal{W},n}^t \langle u_{\mathcal{A},n}|\rho(0)\rangle |u_{\mathcal{A},n}\rangle. \tag{52}$$

For undirected networks, $\mathcal{A}$ is symmetric and its eigenvalues are real. Hence, the viruses die out if the largest eigenvalue $\Lambda_{\mathcal{W},1}$ of $\mathcal{W}$ is less than one, as for all $i$ $\rho_i(t) \to 0$ when $t \to \infty$. Using the expression for $\Lambda_{\mathcal{W},n}$ in Eq. (50), the condition of $\Lambda_{\mathcal{W},1} < 1$ leads to the epidemic threshold

$$\lambda_c < \frac{1}{\Lambda_{\mathcal{A},1}}, \tag{53}$$

which relates $\lambda_c$ to the largest eigenvalue of $\mathcal{A}$.

Other than the steady state, the propagator matrix also allows one to examine the dynamics, as the relaxation to the steady state is dominated by the relaxation mode with the largest eigenvalue when $t \to \infty$. When all eigenvalues of $\mathcal{W}$ are less than 1, the dynamics of the system can be approximated by

$$|\rho(t)\rangle \approx \Lambda_{\mathcal{W},1}^t \langle u_{\mathcal{A},1}|\rho(0)\rangle |u_{\mathcal{A},1}\rangle \qquad \text{as } t \to \infty, \tag{54}$$

which implies that all elements in $|\rho(t)\rangle$, and thus the virus prevalence fraction (the sum of elements in $|\rho(t)\rangle$) decays exponentially with a rate proportional to $\ln \Lambda_{\mathcal{W},1} = \ln(1 - \delta - \nu \Lambda_{\mathcal{A},n})$.

The expression for epidemic threshold (53) is valid for both correlated and uncorrelated networks and provides thresholds for different type of graphs when $\Lambda_{\mathcal{A},1}$ is known. For homogeneous networks or ER networks, $\Lambda_{\mathcal{A},1} \approx \langle k \rangle$ [98] leading to $\lambda_c = 1/\langle k \rangle$ as derived by the probabilistic approach in Eq. (39). For uncorrelated scale-free networks, the expression of $\Lambda_{\mathcal{A},1}$ gives rises to the following epidemic threshold [96],

$$\lambda_c^{\text{SIS,SF}} = \begin{cases} \dfrac{1}{\sqrt{[k_{\max}]}}, & \text{for } \gamma > 2.5 \\ \dfrac{\langle k \rangle}{\langle k^2 \rangle}, & \text{for } 2 < \gamma < 2.5 \end{cases} \tag{55}$$

where $[k_{\max}]$ is the ensemble average of $k_{\max}$ over different realizations of graphs. Equation (55) deviates from Eq. (42) for networks with $\gamma > 2.5$, which implies that $\lambda_c$ vanishes in the thermodynamic limit for all scale-free networks, as supported by $[k_{\max}] \propto N^{1/2}$ for $\gamma \leq 3$ and $[k_{\max}] \propto N^{1/(\gamma-1)}$ for $\gamma > 3$ [99]. It has been argued [96] that the origin of the deviation comes from the assumption of mean-field connectivity in Eq. (40), where the fixed set of specific neighbors of each node (i.e., the quenched topology) is not considered. Further evidence for the infection decay mechanism is obtained by considering a star-like graph, such that $\lambda_c = 1/\sqrt{k_{\max}}$ given by the steady state of Eq. (40), implying the central hub acts as a self-sustained source of infection. This suggests an anti-virus protocol should be based on targeted immunization of networking hubs, which may be more effective than uniform immunization on entire network [100]. Nevertheless, Eq. (55) suggests that the epidemic threshold vanishes in infinite scale-free networks [97, 96].

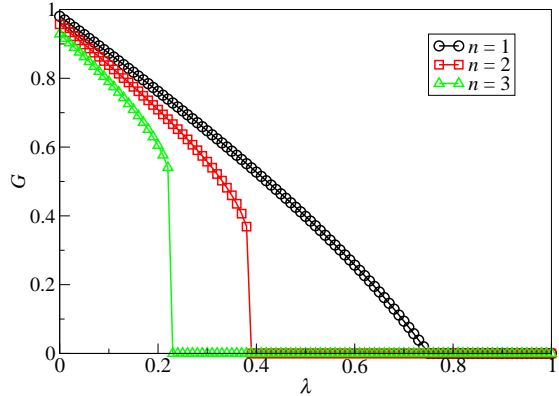

Figure 3: The fraction $C$ of nodes in the giant component after the cascading failures triggered by a fraction $\lambda$ of initially failed nodes. The black curve with circles corresponds to failures on a single ER network with $\langle k \rangle = 4$ and the red and green curves correspond to a NON (network of networks) system with $n = 2$ and $3$ interdependent ER networks arranged in tree structure, obtained by Eq. (62) (see [105] for details).

### 3.3. Cascading Failures

Another dynamical macroscopic behavior of a similar flavor to epidemic spreading is cascading failures. A simple and common example of cascading failures is a massive electric outage, such as the blackout in the Northeast US on $14^{\text{th}}$ August 2003 [101]. The incident was triggered by an initial shut down of a generating plant in Ohio, causing the original loads of the failed plant to be shared among the neighboring plants, overloading some of the transmission lines. These failed power lines caused overloading of other transmission lines, and the process repeated leading to a cascade of failures. Similar phenomenon is also observed and studied in other networking applications such as transportation networks and models of the Internet [102, 103, 104].

To understand the phenomenon, one has to quantify cascading failures. If we consider failed nodes and links as to be removed from the network, a common measure to quantify the influence of failures is the size of the remaining giant component (i.e. the largest connected cluster). Obivously, the smaller the giant component, the more catastrophic the failure. We denote the fraction of nodes in the giant component as $C$, and studies have shown that $C$ is dependent on factors such as the network topology, the centrality and fraction $\lambda$ of the initially failed nodes. As shown by the black circle curve in Fig. 3, $C$ decreases as $\lambda$ increases and exhibits a second-order-like phase transition when $\lambda$ is greater than some threshold value, beyond which no giant component exists after the cascade of failures. This phenomenon occurs in single networks and is already reviewed by Boccaletti et al [19]. Here we are going to review another interesting phenomenon in cascading failures which occur in a *network of networks* (NON) [106, 107, 105] such that the functionality of nodes in one network depend on nodes from other networks, i.e. the failure of nodes in one network trigger the failure of corresponding nodes in the other networks.

*3.3.1. Failures in Systems of Interdependent Networks*

Many networking applications are indeed reliant on such interdependent networks. For instance, the routers of the Internet rely on a power supply network, which in turn depends on the control and monitor through the Internet. Initial failure of nodes in one network may trigger the massive failures in other interdependent networks. Such interdependence between power network and the Internet was the cause of the electricity blackout in Italy is September 2003 [106, 108]. Another interesting example is transportation networks, for instance, the train network is interdependent on the airport network such that a closure of either an airport or the corresponding train station will have a significant influence on its counterpart, and may lead to a cascade of failures on both networks.

Here we will follow the line of [105, 109] to demonstrate the cascading failures in a simple case where individual ER networks are interlinked in a tree-link structure, forming a NON. We will label the $n$ individual networks by Greek letters such that $\alpha = 1, \ldots, n$. For simplicity, we will assume that all the ER networks have the same number of nodes and average degree $\langle k \rangle$. If two networks $\alpha$ and $\beta$ are interdependent, each node in $\alpha$ depends on (at least) one node in $\beta$ and vice versa. In this case, a node in $\alpha$ connected to a failed node in $\beta$ is assumed to fail and is removed from $\alpha$. These failed nodes may fragment network $\alpha$ into isolated sub-clusters. We then assume that only nodes in the giant component of $\alpha$ remains functional, leading to further increase of failed nodes in $\alpha$ which in turn influence nodes in network $\beta$ and other interlinked networks. The process continuous until a stationary state is reached. For a generalization of the above scenario, we refer readers to [105, 109].

To analyze cascading failure in interconnected networks, one denotes $x_\alpha$ to be the fraction of functioning nodes remaining in *all* isolated clusters of network $\alpha$, and $g_\alpha(x_\alpha)$ to be fraction of active nodes belonging to the giant component of $\alpha$. In other words, the fraction of nodes in $\alpha$ that are active and belong to the giant component of is $C_\alpha = x_\alpha g_\alpha(x_\alpha)$ [105]. To obtain an expression for $g_\alpha(x_\alpha)$, one has to employ the generating functions $G_\alpha(z)$, for the degree distribution, and $H_\alpha(z)$ for the branching process, whereby an outgoing link of any node has a probability $k\,P_\alpha(k)/\langle k \rangle_\alpha$ of being connected to a node of degree $k$, which in turn has $k - 1$ outgoing links

$$G_\alpha(z) = \sum_{k=0}^{\infty} z^k \, p_\alpha(k), \tag{56}$$

$$H_\alpha(z) = \sum_{k=0}^{\infty} \frac{k \, P_\alpha(k) \, z^{k-1}}{\langle k \rangle_\alpha} = \frac{G'_\alpha(z)}{G'_\alpha(1)} = \sum_{k=0}^{\infty} z^k q_\alpha(k) \, . \tag{57}$$

These have been introduced in [110] and relate to the degree distribution $p_\alpha(k)$ and excess degree distribution $q_\alpha(k)$ of network $\alpha$ (see Eq. (4)). We note that for ER networks, $p(k) = q(k)$ and hence $G(z) = H(z) = e^{\langle k \rangle(z-1)}$.

Suppose that for the original network $f_\alpha$ denotes the probability that a randomly selected link in $\alpha$ does not lead to the giant component in $\alpha$, then if a link leads to a node with $k - 1$ outgoing links the probability these do not lead to the giant component is $f_\alpha^{k-1}$. Note that this probability is also calculated by $H_\alpha(f_\alpha)$ leading to the recursive relation $f_\alpha = H_\alpha(f_\alpha)$. The probability of a node with degree $k$ not to belong to the giant component is $f_\alpha^k$ and thus the probability that a randomly selected node does belong to the giant component is $g_\alpha = 1 - G_\alpha(f_\alpha)$.

In the damaged network with only a fraction $x_\alpha$ of nodes remaining, $g_\alpha(x_\alpha)$ and $f_\alpha(x_\alpha)$ are obtained by replacing the argument $f_\alpha$ in $H_\alpha(f_\alpha)$ and $G_\alpha(f_\alpha)$ by the expression derived for the remaining nodes $x_\alpha f_\alpha(x_\alpha) + 1 - x_\alpha$, as suggested in [105, 109] based on the results obtained for the branching process studied in [111]. These give rise to

$$g_\alpha(x_\alpha) = 1 - G_\alpha[x_\alpha f_\alpha(x_\alpha) + 1 - x_\alpha], \tag{58}$$
$$f_\alpha(x_\alpha) = H_\alpha[x_\alpha f_\alpha(x_\alpha) + 1 - x_\alpha]. \tag{59}$$

For ER networks with average degree $\langle k \rangle_\alpha$, since $G(z) = H(z)$, the relation between the two functions takes a particularly simple form

$$g_\alpha(x_\alpha) = 1 - f_\alpha(x_\alpha) = 1 - e^{\langle k \rangle_\alpha x_\alpha (f_\alpha - 1)} = 1 - e^{-\langle k \rangle_\alpha x_\alpha g_\alpha(x_\alpha)}. \tag{60}$$

For general topologies it has been shown [105] that

$$C = \prod_{\alpha=1}^{n} (1 - \lambda_\alpha) g_\alpha(x_\alpha) \tag{61}$$

where $C$ is the *fraction* of nodes in the mutual giant component among all networks. In cases where a one-to-one correspondence of nodes is established between all pairs of interdependent networks, and where $C_\alpha$ denotes the *fraction* of nodes in the giant component of network $\alpha$, then $C_\alpha = C$ for all $\alpha$ since the remaining functional nodes in pairs of interdependent network must be connected to each other. The parameter $\lambda_\alpha$ is the fraction of initially failed nodes in network $\alpha$. Assume initial failures occur only in one network, i.e. $\lambda_\beta = \lambda$ and $\lambda_\alpha = 0$ for all $\alpha \neq \beta$, then by substitution of Eq. (60) and the relations $C = x_\alpha g_\alpha(x_\alpha)$ for all $\alpha$ into the Eq. (61), one obtains a self-consistent equation in $C$ [105, 109] given by

$$C = (1 - \lambda) \left[1 - e^{-\langle k \rangle C}\right]^n \tag{62}$$

where the assumption $\langle k \rangle_\alpha = \langle k \rangle$ has been utilized.

The largest solution of $C$ in Eq. (62) is shown in Fig. 3 as a function of $\lambda$. As we can see, while the giant component $C$ exhibits a second-order-like phase transition and vanishes when $\lambda$ is greater than a threshold value when $n = 1$; a first-order-like phase transition is observed when $n > 1$ suggesting a different cascading failure phenomenon occurs in NON compared to single networks. The physical reasons behind the difference are investigated in [112]. Moreover, these results also suggest that NON is more fragile to cascading failures given the same fraction of initially failed nodes, since the giant component is smaller and vanishes at smaller for non-zero $\lambda$.

However, network interdependence is not necessarily disadvantageous. A study of sandpile models on interdependent networks reveals that large cascades are mitigated by a small fraction of inter-links between networks [113]. This may be relevant to the tolerance of congestion in transportation and communication networks [103], where load balancing between networks such as metro and bus networks may suppress congestion cascades. As most of these studies are specific to particular networks or architectures, we refer the reader to the original publications for further details.

## 4. Networking as Disordered Systems

One area of statistical physics that is particularly relevant to networking is the study of *disordered systems*, where interactions between variables are fixed, or *quenched*, and do not evolve in time, in contrast to the *dynamical* system variables. If we consider for instance, a random walk in a particular network, the network topology is quenched (fixed) unlike the dynamical on-site probability of the walker. We note that most networking systems are disordered, as networks generally have different quenched topology even if they are characterized by the same degree distribution $p(k)$. The seemingly small difference in the topology may lead to severe deviations in their behavior; for example the results of epidemic threshold in scale-free networks [96]. Thus to obtain the typical behaviors of networking systems, one has to employ techniques which properly consider the microscopic properties of specific heterogeneous topologies through averaging over network instances. In this section, we will see that various networking-related problems are solved by techniques developed in the field of disordered systems.

In statistical physics, the most extensively studied disordered system is spin glass, a magnet with randomly distributed ferromagnetic and anti-ferromagnetic interaction couplings between spin variables. The generic Hamiltonian of spin glass can be written as

$$H = -\sum_{(ij)} a_{ij} J_{ij} s_i s_j, \tag{63}$$

where $a_{ij}$ is an element of the adjacency matrix, $s_i$ denotes the spin-state of node $i$ and the set $J_{ij}$ are couplings between spins, which constitute the quenched disorder in the system. When $J_{ij} = J$ for all pairs of $i$ and $j$, the spin glass reduces to a ferromagnet. In contrast to a ferromagnet with only two ground states at zero temperature, spin glasses with randomly selected sets of $J_{ij}$ generally have different ground states and give rise to a very rich energy landscape. Technically, the coupling distributions make the analysis, and averaging over the disorder in particular, difficult.

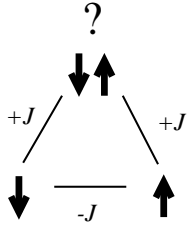

Figure 4: An example of frustrated system with 3 spins. $+J$ and $-J$ corresponds to ferromagnetic and anti-ferromagnetic interactions of strength $J$.

Spin glasses were first studied on a regular lattice in the Edwards-Anderson (EA) model [114] and on a fully connected network in the Sherrington-Kirkpatrick (SK) model [115]. They were later extended to ER networks [116], scale-free networks [117] and small world networks [118].

To obtain the typical behaviors of disordered systems, one can make use of their *self-averaging* property, i.e., in the thermodynamic limit the free energy per degree of freedom (i.e. $\mathcal{F}/N$) of a particular system sampled from a given disorder distribution is equal to the average over systems with different disorders sampled from the same distribution. Away from the phase transition point, self-averaging can be considered as a consequence of the central limit theorem since the free energy is an extensive quantity and increases linearly with the system size; the distribution of its value becomes more sharply peaked as $N \to \infty$. Thus, the typical properties of disordered systems are usually represented by their *ensemble average* in the thermodynamic limit. However, at the phase transition point, self-averaging is not necessarily valid due to the presence of long-range correlations.

Apart from the technical difficulties in disorder averaging, *frustration* between individual variables also complicate the state space of such systems. Frustrations occur when individual optimality cannot be achieved globally and some variables are found in sub-optimal states. A simple example of frustrated system is given in Fig. 4, where one of the three spin interaction is always in a sub-optimal state regardless of the state of the top spin. When a finite fraction of frustrated variables are present, the system has to overcome energy barriers to shift from one low lying state to another. This creates an energy landscape with numerous valleys, and leads to non-ergodicity and difficulties in locating the optimal state.

*4.1. Frequency Allocation*

There are many disordered networking problems which can be analyzed using techniques from spin glass theory, among them is the problem of frequency allocation in radio and television broadcasting as well as Wireless Local Area Network (WLAN) access points [119]. An example of frequency allocation problem is shown in Fig. 5(a) where radio broadcasting to an area is achieved by 4 broadcasting stations (triangles). If all the stations adopt the same broadcasting radio frequency, areas lying in the intersection of individual coverages may suffer from interference. Individual stations thus adopt different frequencies as shown in Fig. 5(b), such that interference is minimized in areas with overlapping reception. The same also applies to WLAN as neighboring access points generally use non-overlapping channels for signal provisions to avoid interference.

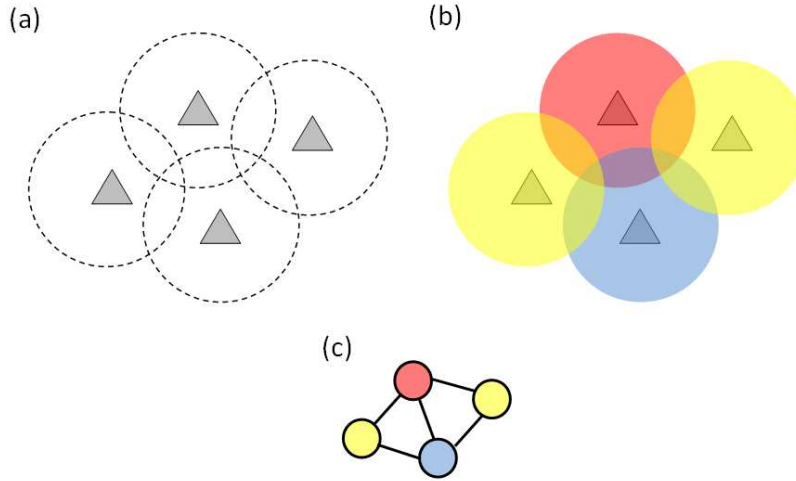

Figure 5: (a) An example of radio broadcasting which shows 4 broadcasting stations (triangles) and their corresponding range of transmission (dashed circles). (b) Frequency allocation with individual colors corresponding to the specific frequencies. (c) Mapping the frequency allocation problem onto the graph coloring problem on a network.

The assignment of frequencies in the example of Fig. 5(b) is equivalent to coloring the graph in Fig. 5(c) where nodes represent stations and edges exist when there is an overlap between transmission areas of neighboring stations. The problem of finding the minimal number of channels for non-interfering communication, can be mapped to the *graph coloring* problem of theoretical computer science, which corresponds to the coloring of nodes using a limited number of colors such that no neighbors share the same color. The minimal number of colors such that the graph is *colorable*, i.e. all neighbors have different colors, is termed the *chromatic number*. In addition, a variant of the graph coloring problem, called the color diversity problem [120], which we will briefly describe here is relevant to distributed file storage when accessibility of *different* file segments is desired, for instance, in peer-to-peer (P2P) networks.

*4.1.1. The Graph Coloring Problem*

Graph coloring on regular lattice and planar graphs has been extensively studied in discrete mathematics. For any planar graphs, Appel and Haken [121, 122] proved that four colors are sufficient. However, the problem of finding whether a number of colors is sufficient is NP-complete and requires an exponentially growing computing time-steps with respect to the system size $N$, for general graphs. While computer scientists study the worst case scenarios and establish upper and lower bounds for the colorable threshold, physicists aim to predict the typical behavior which involves averaging samples of network topologies that correspond to the disorder.

To formulate the problem in the framework of statistical physics, we denote $\sigma_i = 1, \cdots, Q$ to be the color of node $i$ and write the Hamiltonian of the system as

$$H = \sum_{(ij)} a_{ij} \delta_{\sigma_i, \sigma_j} = \sum_{(ij)} a_{ij} \sum_{q=1}^{Q} \delta_{\sigma_i, q} \delta_{\sigma_j, q}, \qquad (64)$$

which is equivalent to the Hamiltonian of a $Q$-state anitferromagnetic Potts model [123]. To obtain the optimal color configuration, we find the state that minimizes the Hamiltonian $H$. If the network is colorable, the ground state energy is zero, and positive otherwise. We thus examine the ground state energy of $H$ to determine whether a network is colorable or not, and find the fraction of edges with forbidden color assignments. Networks with more edges contain more constraints and a larger number of colors is needed to satisfy them. Given a network model with tunable degree distribution and available colors $Q$, we expect to see a transition at $\langle k \rangle = \langle k \rangle_c$ below which the graph is colorable and above which it is not. We call $\langle k \rangle_c$ the *coloring threshold* of the graph.

Here we outline two general techniques in spin glass theory, namely the replica and cavity approaches [37, 38], which can be used to analyze the graph coloring problem as well as other disordered systems. Thus the present subsection serves as a description of the graph coloring problem as well as an example of how spin glass theory can be applied to other networking problems.

***The Replica Method.*** To derive the typical behavior of a disordered system, we have to evaluate the disorder average of the free energy $[F]$, where $[\cdots]$ denotes the average over the quenched disorder. As $F = -T \ln Z$, where $T$ and $Z$ represent the temperature and the partition function, respectively; the average free energy $[F]$ can be obtained by averaging a logarithmic function which is technically difficult. One thus employs the replica trick

$$[\ln Z] = \lim_{n \to 0} \frac{[Z^n] - 1}{n} \qquad (65)$$

to evaluate the disorder average of $Z^n$ instead of $\ln Z$.

As replica calculations are generally rather involved, we will merely outline here the main steps and refer the interested reader to specific literature on the replica method in general [37, 38, 39] and on the graph coloring problem in particular [124].

To apply the replica trick one first introduces the replicated partition function $Z^n$ averaged over the disorder

$$[Z^n] = \operatorname*{Tr}_{\mathcal{A}} \rho(\mathcal{A}) \left( \operatorname*{Tr}_{\sigma_1 \cdots \sigma_N} e^{-\beta \sum_{(ij)} a_{ij} \delta_{\sigma_i, \sigma_j}} \right)^n = \operatorname*{Tr}_{\mathcal{A}} \rho(\mathcal{A}) \operatorname*{Tr}_{\vec{\sigma}_1 \cdots \vec{\sigma}_N} \prod_{\alpha=1}^{n} e^{-\beta \sum_{(ij)} a_{ij} \delta_{\sigma_i^\alpha, \sigma_j^\alpha}} \qquad (66)$$

where the trace represents a summation over all the possible values of the corresponding variable, $\rho(\mathcal{A})$ corresponds to the probability of generating a graph with adjacency matrix $\mathcal{A}$, while $\vec{\sigma}_i$ corresponds to the replicated vector $\sigma_i^\alpha$ with replica indices $\alpha = 1, \cdots, n$. As one can see, the disorder average over the topology, represented by $\mathbf{Tr}_\mathcal{A}$, can be readily carried out for a given ensemble distribution $\rho(\mathcal{A})$. For example, $\rho(\mathcal{A})$ in ER networks can be represented by the product $\prod_{(ij)}[(1-p)\delta_{a_{ij},0} + p\delta_{a_{ij},1}]$ such that after the disorder average $[Z^n]$ becomes

$$[Z^n] = \mathbf{Tr}_{\vec{\sigma}_1 \cdots \vec{\sigma}_N} \prod_{(ij)} \left(1 - p + p \prod_\alpha e^{-\beta \delta_{\sigma_i^\alpha, \sigma_j^\alpha}}\right) \approx \mathbf{Tr}_{\vec{\sigma}_1 \cdots \vec{\sigma}_N} \prod_{(ij)} \exp\left[p \prod_\alpha e^{-\beta \delta_{\sigma_i^\alpha, \sigma_j^\alpha}} - p\right], \quad (67)$$

the assumption of sparse connectivity, i.e. a small value of $p$, facilitates the derivation of the final expression.

To proceed, we then decouple the interaction between $i$ and $j$ by rewriting

$$e^{-\beta \delta_{\sigma_i^\alpha, \sigma_j^\alpha}} = e^{-\beta \sum_{q=1}^Q \delta_{q,\sigma_i^\alpha} \delta_{q,\sigma_j^\alpha}} = 1 - (1 - e^{-\beta}) \sum_{q=1}^Q \delta_{q,\sigma_i^\alpha} \delta_{q,\sigma_j^\alpha} \quad (68)$$

and expand this expression over the product of replica indices $\alpha$, which leads to

$$[Z^n] = \mathbf{Tr}_{\vec{\sigma}_1 \cdots \vec{\sigma}_N} \exp\left[p \left(\sum_{m=0}^n (1 - e^{-\beta})^m \sum_{(\alpha)_m} \sum_{\{q\}_m} \sum_{(ij)} \prod_{l=1}^m (\delta_{q_l, \sigma_i^{\alpha_l}} \delta_{q_l, \sigma_j^{\alpha_l}})\right)\right] e^{\frac{pN(N-1)}{2}}, \quad (69)$$

where we denote $(\alpha)_m$ as the set of ordered indices $(\alpha_1, \cdots, \alpha_m)$, the notation $(ij)$ is used for an ordered set of indices and $\{q\}_m$ is the set of colors $\{q_1, \cdots, q_m\}$. We note that in Eq. (69), the term $\sum_{(ij)} \prod_{l=1}^m (\delta_{q_l, \sigma_i^{\alpha_l}} \delta_{q_l, \sigma_j^{\alpha_l}})$ can be written as $(\sum_i \prod_{l=1}^m \delta_{q_l, \sigma_i^{\alpha_l}})^2/2$ by neglecting lower order terms of $O(N)$. The interaction between $i$ and $j$ is decoupled, by further defining, for each value of $m$ and corresponding set of $(\alpha)_m$ and $\{q\}_m$, the macroscopic order parameters $q_{\{q\}_m}^{(\alpha)_m} = \sum_i \prod_{l=1}^m \delta_{q_l, \sigma_i^{\alpha_l}}$.

In general, to make further progress in the derivation one has to make further assumptions about the correlation between replicas, such as the *replica symmetry* (RS) assumption that replica indices are symmetric and indistinguishable. Once the order parameters are represented in parametric form and the limit of $n \to 0$ is taken, $[Z^n]$ can be evaluated by the method of steepest descent as $N \to \infty$ and the expressions for free energy, energy and entropy can be derived.

To determine the colorable threshold $\langle k \rangle_c$ for the ER networks, one evaluates the ground state energy as a function $p$ under the RS assumption to find that $\langle k \rangle_c = 1$ for $Q = 2$, which coincides with simulation results [124]. The phase diagram obtained is physical, such that all phases have non-negative energy and entropy. The RS assumption appears to break down for higher $Q$ values close to the threshold; this is manifested by the emergence of negative entropy values. We will discuss an improved assumption termed the *replica symmetric breaking* (RSB) ansatz in the next subsection and in Section 5.2, which facilitated the derivation of an improved threshold $\langle k \rangle_c \approx 5.1$ for the case of $Q = 3$ and similar results for different $Q$ values and degree distributions [125, 126].

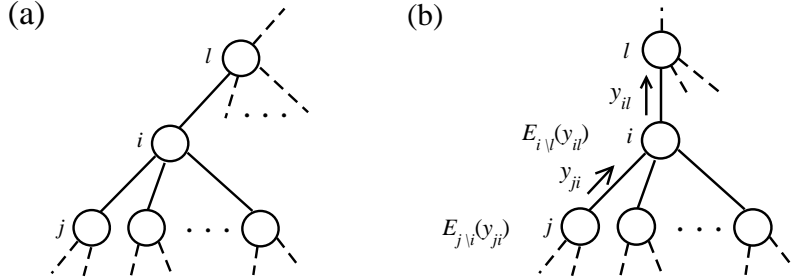

Figure 6: The locally tree-like structure in networks.

***The Cavity Method.*** An alternative method to carry out averages over quenched disorder is the cavity method based on calculating field distributions once an individual vertex/edge has been removed. The cavity approach is particularly useful in cases of a locally tree like structure as shown in Fig. 6(a). In these cases it gives rise to efficient algorithms for specific instances in addition to insightful analytical results.

To solve the graph coloring problem by the cavity approach, we write the energy of the *descendant* node $j$ in the absence of *ancestor* node $i$,

$$E_{j\backslash i}(\sigma_j) = C_{j\backslash i} + \sum_{q=1}^{Q} h_{j\backslash i}^q \delta_{q,\sigma_j}, \tag{70}$$

by introducing the *cavity field* $\vec{h}_{j\backslash i} = (h_{j\backslash i}^1, \cdots, h_{j\backslash i}^Q)$ [127, 125] and a constant $C_{j\backslash i}$, with superscripts denoting the color $q = 1, \cdots, Q$ and subscripts $j\backslash i$ denoting node $j$ in the absence of $i$. The most favorable color would have the smallest field. The *cavity energy function* $E_{j\backslash i}(\sigma_j)$ describes the energy of the sub-tree terminated at $j$ in the absence of $i$. In the zero temperature limit, it is expressed in terms of the cavity energy functions of its descendants $E_{j\backslash i}(\sigma_j)$

$$E_{i\backslash l}(\sigma_i) = \min_{\vec{\sigma}} \left[ \sum_{j \neq l} a_{ij} \left( E_{j\backslash i}(\sigma_j) + \delta_{\sigma_i,\sigma_j} \right) \right] \tag{71}$$

in which one assumes that the descendant $j$'s are independent in the absence of $i$, i.e. the network is of a locally tree-like structure at $i$. Though this assumption does not hold in finite networks with loops, algorithms based on the cavity approach have negligible errors in finite size systems with low connectivity [128, 129]. To make Eq. (71) into an iterative form one can rewrite $E_{i\backslash l}(\sigma_i)$ as

$$E_{i\backslash l}(\sigma_i) = C_{i\backslash l} + \sum_{q=1}^{Q} h_{i\backslash l,q} \delta_{q,\sigma_i}, \tag{72}$$

such that $\vec{h}_{i\backslash l}$ and $C_{i\backslash l}$ are given by

$$h_{i\backslash l}^q = \sum_{j\neq l} a_{ij}\delta_{q,q_{j\backslash i}^*}, \tag{73}$$

$$C_{i\backslash l} = \sum_{j\neq l} a_{ij}(C_{j\backslash i} + h_{j\backslash i}^*) \tag{74}$$

where

$$q_{j\backslash i}^* = \begin{cases} \arg_q \min(h_{j\backslash i}^1, \cdots, h_{j\backslash i}^Q) & \text{if a unique optimal color exists for } j \\ -1 & \text{if no unique optimal color exists for } j \end{cases}, \tag{75}$$

$$h_{j\backslash i}^* = \min(h_{j\backslash i}^1, \cdots, h_{j\backslash i}^Q). \tag{76}$$

Physically, Eq. (73) implies that if $i$ selects the color that coincide with the unique optimal color of $j$, the cavity field for that color increases by one.

Note that Eq. (73) takes an iterative form of the cavity fields, which leads to potential algorithmic applications as we will see in Section 4.2. To obtain the macroscopic properties of the system, we first evaluate the distribution $P(\vec{h})$ of cavity fields by solving the self consistent equation [127, 125, 126]

$$P(\vec{h}) = \sum_{z=0}^{\infty} q(z) \int \prod_{j=1}^{z} d\vec{h}_j P(\vec{h}_j) \prod_{q=1}^{Q} \delta\left[h^q - \sum_j \delta_{q,q_j^*(\vec{h}_j)}\right], \tag{77}$$

where $q(z)$ is the *excess degree* distribution of $z = k - 1$, the remaining degree of a node arrived from a random link. Excess degree is used since the cavity field is obtained by removing one edge thus effectively reducing the number of neighbors. We have emphasized the dependence of $q_j^*$ on $\vec{h}_j$ by writing the latter as the argument of $q_j^*$. The above self consistent equation can be solved by population dynamics [128], where the field distribution $P(\vec{h})$ is modeled by an evolving population of variables; the fields of individual variables are calculated iteratively via Eq. (77) by sampling from $P(\vec{h})$.

Finally, we can calculate the ground state energy of the system by expressing the energy change of an additional node and an additional edge as

$$\Delta E_{\text{node}} = \min_{\sigma_i,\vec{\sigma}}\left[\sum_j a_{ij}\left(E_{j\backslash i}(\sigma_j) + \delta_{\sigma_i,\sigma_j}\right)\right] - \sum_j a_{ij} \min_{\sigma_j}\left[E_{j\backslash i}(\sigma_j)\right], \tag{78}$$

$$\Delta E_{\text{link}} = \min_{\sigma_i,\sigma_j}\left[E_{i\backslash j}(\sigma_i) + E_{j\backslash i}(\sigma_j) + \delta_{\sigma_i,\sigma_j}\right] - \min_{\sigma_i}\left[E_{i\backslash j}(\sigma_i)\right] - \min_{\sigma_j}\left[E_{j\backslash i}(\sigma_j)\right]. \tag{79}$$

Using similar arguments as in the derivation of Eq. (73), one can show that $\Delta E_{\text{node}} = \min_{\sigma_i}(\sum_j a_{ij}\delta_{\sigma_i,q_{j\backslash i}^*})$ and $\Delta E_{\text{link}} = \min_{\sigma_i}(h_{i\backslash j}^{\sigma_i} + \delta_{\sigma_i,q_{j\backslash i}^*}) - h_{i\backslash j}^*$, such that the average $\Delta E_{\text{node}}$

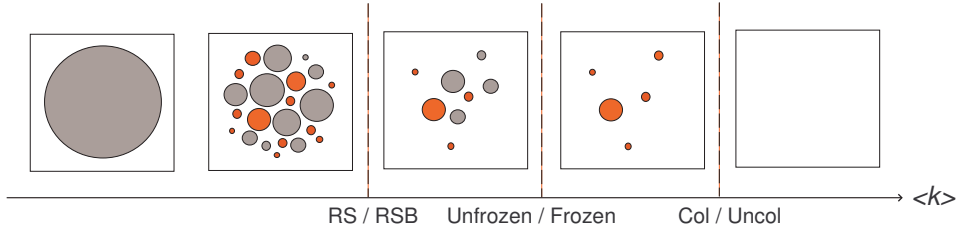

Figure 7: The schematic phase diagram obtained by Krzakala et al [126, 130, 131] which indicates the changes in solution space of the graph coloring problem as a function of $\langle k \rangle$ (given $Q$ colors). Solution clusters are represented by grey circles, and red circles correspond to clusters with frozen variables.

and $\Delta E_{\text{link}}$ can be expressed in terms of $P(\vec{h})$ as

$$\langle \Delta E_{\text{node}} \rangle = \sum_{k=0}^{\infty} p(k) \int \prod_{j=1}^{k} d\vec{h}_j P(\vec{h}_j) \min_q \left( \sum_j \delta_{q,q_j^*} \right), \quad (80)$$

$$\langle \Delta E_{\text{link}} \rangle = \int d\vec{h}_i d\vec{h}_j P(\vec{h}_i) P(\vec{h}_j) \left[ \min_{q_i} (h_i^{q_i} + \delta_{q_i,q_j^*}) - h_i^* \right]. \quad (81)$$

The resulting average energy $\langle E \rangle$ per node is then given by $\langle \Delta E_{\text{node}} \rangle - (\langle \Delta E_{\text{link}} \rangle \langle k \rangle)/2$, where the second term compensates the energy contribution from additional edges when a node is added to the network, to maintain the ratio between nodes and edges.

So far we have assumed that a single $P(\vec{h})$ is sufficient to describe the system, which is equivalent to the replica symmetry assumption in the replica approach. In the RS assumption, a single cluster of states dominates the phase space and a single $P(\vec{h})$ is sufficient to describe it. It was shown that the cavity calculations with the RS assumption predict the existence of an unphysical phase characterized by negative energy [127, 125]. Under an improved assumption called the one-step replica symmetry breaking ansatz (1RSB) [37, 38, 39], a distribution of $P(\vec{h})$ is used to describe the system such that the derived energy is non-negative for all phases. Physically, the 1RSB ansatz assumes that the solutions are grouped in clusters in the solution space, and different $P(\vec{h})$ describe different clusters.

If we consider solutions to be neighbors if they differ in only one color assignment, a phase diagram indicating the changes in solution space as a function of $\langle k \rangle$ in ER systems was obtained by Krzakala et al [126, 130, 131] by applying the 1RSB ansatz of the cavity approach, as shown in Fig. 7. Here solution clusters are represented by grey circles, and red circles correspond to clusters with frozen variables, i.e. variables which exhibit the same value in all solutions of the clusters. As we can see from the first two panels, the single solution clusters segments into clusters as $\langle k \rangle$ increases. When connectivity further increases, the problem becomes harder to solve as the number of solution clusters greatly decreases; the system transits to a regime where the RSB ansatz is valid. In this phase, numerous low-lying states exist which hinders local search algorithms from finding optimally colorable solutions [127, 125]. The color assignment becomes even more difficult when $\langle k \rangle$ increases further and the solution space is dominated by frozen clusters. Finally, beyond the colorable threshold no solution exists. This phase diagram is important to most combinatorial problems, including problems relevant to the study of networking and especially those described in the subsequent subsections. More discussions on the 1RSB ansatz are found in Section 5.2.

*4.1.2. The Color Diversity Problem*

A variant of the graph coloring problem, the color diversity problem, is relevant to distributed file storage particularly in overlay networks. In this problem, we search for the color configuration that maximizes the number of different colors in the nearest neighborhood of every node. The reasons become obvious if we consider different colors as different file segments and nodes can retrieve the complete file by searching only in their immediate neighborhoods [120]. To achieve the goal, we search for the variable state that minimizes the Hamiltonian

$$H = \sum_i \left( \sum_j a_{ij} \delta_{\sigma_i, \sigma_j} + \sum_{j,k} a_{ij} a_{ik} \delta_{\sigma_j, \sigma_k} \right) = \sum_i \sum_{q=1}^Q \left[ \delta_{\sigma_i, q} + \sum_j a_{ij} \delta_{q, \sigma_j} \right]^2, \qquad (82)$$

where $H$ increases with the number of identical colors in the nearest neighborhood of a node. Although the above Hamiltonian maximizes the color diversity at each neighborhood, it does not guarantee a complete set of color retrieval. To maximize the fraction of nodes with a complete set of colors, an alternative Hamiltonian can be formulated

$$H = -\sum_i \Theta \left[ Q - \sum_{q=1}^Q \Theta \left( \delta_{\sigma_i, q} + \sum_j a_{ij} \delta_{q, \sigma_j} \right) \right], \qquad (83)$$

where the heaviside function $\Theta(x) = 1$ when $x > 0$ and $\Theta(x) = 0$ otherwise. By minimizing the Hamiltonian (83), the fraction of nodes with a complete set of colors is maximized, but without guaranteeing the uniformity of the color distribution.

Unlike the graph coloring problem on random graphs, which becomes more difficult as $\langle k \rangle$ increases, the color diversity problem becomes harder when $\langle k \rangle$ decreases, as there are less neighbors, hence less sources, to retrieve a complete set of color. Given $Q$ colors and if we denote the number of nodes with incomplete set of colors to be $f$, we expect to see a phase transition from a region of $f = 0$ to $f > 0$ when $\langle k \rangle$ becomes smaller than a threshold value $\langle k \rangle_c$, resembling the transition from colorable region to uncolorable region in graph coloring. Such phenomenon is studied by the cavity approach in [120, 132]. In the context of file retrieval, search beyond the nearest neighbors is required to recover the complete file when $\langle k \rangle < \langle k \rangle_c$.

*4.2. Resource Allocation*

Here we describe another disordered system, that gives rise to the resource allocation problem, which is highly relevant to networking applications such as the distribution of load in computer clusters. In addition, a variant of the resource allocation problem can be mapped onto a path optimization scenario where sensors communicate with a central base station through routes on a sparse graph; it may also play a role in exploring instabilities and redundancies in communication networks. Another variant of the problem that is relevant to routing in sensor and ad-hoc networks is the source location problem aimed at identifying the best allocation of base-stations/routers in order to optimize a given cost while carrying the communication tasks at hand. some aspects of the ubiquitous problem of routing may also be tackled using approaches developed for resource allocation as we will see later on.

*4.2.1. Re-distribution of Resources*

In the resource re-distribution problem, one consider each node $i$ to have resource $\lambda_i$, randomly drawn from a distribution $\rho(\lambda)$, such that $\lambda > 0$ and $\lambda < 0$ correspond to a surplus and a shortage of resource. Resources are then re-allocated between the nodes. The real variable $y_{ji}$ represents the flow of resources from $j$ to $i$, the final resource on $i$ is given by $r_i = \lambda_i + \sum_{j \in \mathcal{L}_i} y_{ji}$, where $\mathcal{L}_i$ corresponds to the set of neighbors of $i$. The randomly sampled set of $\lambda_i$ adds another level of disorder on top of the network topology. One then minimizes the Hamiltonian given by

$$H = \alpha \sum_i \psi(r_i) + \sum_{(ij)} a_{ij} \phi(y_{ij}), \tag{84}$$

where $\psi(r)$, the *shortage cost* which penalizes nodes with negative $r$, and $\phi(y)$, the *transportation cost*, for instance, to power consumption of electric current, take the form of

$$\psi(r) = \Theta(-r), \tag{85}$$
$$\phi(y) = \frac{y^2}{2}. \tag{86}$$

The heaviside step function takes the value $\Theta(x) = 1$ for $x > 0$ and 0 otherwise. The coefficient $\alpha$ in Eq. (84) acts as a parameter which controls the relative weight between the two costs.

When $\sum_i \lambda_i \geq 0$ and $\alpha \to \infty$, the ground state of $H$ corresponds to a state with $r_i \geq 0$, $\forall i$, subject to the minimal transportation cost. This case corresponds to the *load balancing* problem [133, 134] if one consider $\lambda$ as the *capacity* of a node, and the minimization of $H$ is equivalent to the balance of loads with the minimal migration of tasks. In the context of networking and computation, these could be considered as computing power and tasks to be carried out. On the other hand, if one consider finite positive $\alpha$, nodes are allowed to have negative final resources provided that $H$ is still minimized. In this case, the problem is equivalent to the *source location* problem [135, 136] where the positions of sources are optimally located in a network to minimize transportation and the local penalty correspond to the cost of providing the resource at the node. Details of the source location problem are discussed in Section 4.2.2.

We first describe how the cavity method is applied to solve the resource allocation problem. To apply the cavity approach, one considers the local tree structure as in Fig. 6(b) and writes the cavity energy function $E_{i\backslash l}(y_{il})$ of a node $i$ with a current $y_{il}$ drawn by its ancestor $l$, as

$$E_{i\backslash l}(y_{il}) = \mathcal{H}(E_{j_1\backslash i}, \cdots, E_{j_{k_i-1}\backslash i}; \lambda_i, y_{il}) \tag{87}$$

where $j_1, \cdots, j_{k_i-1}$ are the $k_i - 1$ neighbors of $i$ other than $l$, and the functional $\mathcal{H}$ corresponds to

$$\mathcal{H}(E_{j_1\backslash i}, \cdots, E_{j_{k_i-1}\backslash i}; \lambda_i, y_{il}) =$$
$$\min_{\{y_{ji}\}} \left[ \sum_{j \in \mathcal{L}_i \backslash l} E_{j\backslash i}(y_{ji}) + \alpha\psi\left(\lambda_i - y_{il} + \sum_{j \in \mathcal{L}_i \backslash l} y_{ji}\right) + \sum_{j \in \mathcal{L}_i \backslash l} \phi(y_{ji}) \right] \tag{88}$$

in the zero-temperature limit. The first term represents the energies of descendants, the second the penalty for negative resource once the flow has been taken into account, and the third is the transportation cost for a given current drawn from $i$. We note that $E_{j\backslash i}(y_{ji})$ is an extensive quantity but depends on the number of iterations; these are calculated recursively from a vertex dependent intensive energy. One first writes $E_{j\backslash i}(y_{ji})$ as a sum of two terms,

$$E_{j\backslash i}(y_{ji}) = E^V_{j\backslash i}(y_{ji}) + E_{j\backslash i}(0), \tag{89}$$

where $E^V_{j\backslash i}(y_{ji})$ is called the *vertex cavity energy* such that $E^V_{j\backslash i}(0) = 0$, i.e., if no current is drawn the vertex cavity energy remains the same. This allows us to rewrite Eq. (87) as a recursion of the intensive quantity $E^V$ [136], given by

$$E^V_{i\backslash l}(y_{il}) = \mathcal{H}(E^V_{j_1\backslash i}, \cdots, E^V_{j_{k_i-1}\backslash i}; \lambda_i, y_{il}) - \mathcal{H}(E^V_{j_1\backslash i}, \cdots, E^V_{j_{k_i-1}\backslash i}; \lambda_i, 0), \tag{90}$$

which corresponds to a self-consistent form of $E^V$ as Eq. (90) satisfies $E^V_{i\backslash l}(0) = 0$. With this recursion of intensive $E^V$, one then obtains the self-consistent equation analogous to Eq. (77) for the functional distribution $P[E^V(y)]$ as

$$P[E^V(y)] = \int d\lambda \rho(\lambda) \sum_{z=0}^{\infty} q(z) \int \prod_{j=1}^{z} dE_j(y_j) P[E^V_j(y_j)]$$
$$\times \delta\left[E^V(y) - \mathcal{H}(E^V_{j_1}, \cdots, E^V_{j_z}; \lambda, y) + \mathcal{H}(E^V_{j_1}, \cdots, E^V_{j_z}; \lambda, 0)\right], \tag{91}$$

where $q(z)$ is the excess degree distribution of Eq. (4). Given a stable solution of $P[E^V(y)]$, the ground state energy of the system is obtained by evaluating [136]

$$\langle \Delta E_{\text{node}} \rangle = \int d\lambda \rho(\lambda) \sum_{k=0}^{\infty} p(k) \int \prod_{j=1}^{k} dE_j(y_j) P[E^V_j(y_j)] \mathcal{H}(E^V_{j_1}, \cdots, E^V_{j_k}; \lambda, 0), \tag{92}$$

$$\langle \Delta E_{\text{link}} \rangle = \int dE_1(y) dE_2(y) \min_y \left[E_1(y) + E_2(-y) + \phi(y)\right], \tag{93}$$

such that the average energy per node is $\langle E \rangle = \langle \Delta E_{\text{node}} \rangle - (\langle \Delta E_{\text{link}} \rangle \langle k \rangle)/2$.

Note that Eqs. (92)-(93) involve solving a functional distribution $P[E^V(y)]$, which is in general infeasible. However, the form of $\phi$ and $\psi$ in Eqs. (86) and (85) results in a piecewise quadratic expression for $E^V$, which greatly reduces the functional space of $P[E^V(y)]$; we refer readers to Refs. [134, 136] for technical details. The complex recursion of $E^V$ turns into a simple message passing algorithm which involves messages with only two real values.

As $\alpha \to \infty$, shortage in the load balancing problem can be expressed as a constraint $\lambda_i - y_{il} + \sum_{j \in \mathcal{L}_i \setminus l} y_{ji} \geq 0$ such that $\mathcal{H}$ in Eq. (88) takes the form of

$$\mathcal{H}(E_{j_1 \setminus i}, \cdots, E_{j_{k_i-1} \setminus i}; \lambda_i, y_{il}) = \min_{\{\{y_{ji}\} | \lambda_i - y_{il} + \sum_{j \in \mathcal{L}_i \setminus l} y_{ji} \geq 0\}} \left[ \sum_{j \in \mathcal{L}_i \setminus l} E_{j \setminus i}(y_{ji}) + \sum_{j \in \mathcal{L}_i \setminus l} \frac{y_{ji}^2}{2} \right]. \quad (94)$$

One can derive an optimization algorithm by explicitly evaluating the above expression through the expansion of $E_{j \setminus i}(y_{ji})$ in the neighborhood of a constant value $\tilde{y}_{ji}$. Assuming a small variation of $\epsilon_{ji}$ from $\tilde{y}_{ji}$, $E_{j \setminus i}(y_{ji})$ is given by

$$E_{j \setminus i}(y_{ji}) = E_{j \setminus i}(\tilde{y}_{ji} + \epsilon_{ji}) \approx E_{j \setminus i}(\tilde{y}_{ji}) + a_{j \setminus i} \epsilon_{ji} + b_{j \setminus i} \frac{\epsilon_{ji}^2}{2}, \quad (95)$$

to the second order in $\epsilon_{ji}$. Evaluating $\mathcal{H}$ in Eq. (94) is equivalent to minimizing the Lagrangian

$$\begin{aligned} L_{i \setminus l} &= \sum_{j \in \mathcal{L}_i \setminus l} \left[ E_{j \setminus i}(\tilde{y}_{ji}) + a_{j \setminus i} \epsilon_{ji} + b_{j \setminus i} \frac{\epsilon_{ji}^2}{2} \right] + \sum_{j \in \mathcal{L}_i \setminus l} \frac{(\tilde{y}_{ji} + \epsilon_{ji})^2}{2} \\ &+ \mu_i \left( \lambda_i - y_{il} + \sum_{j \in \mathcal{L}_i \setminus l} (\tilde{y}_{ji} + \epsilon_{ji}) \right), \end{aligned} \quad (96)$$

with respect to $\epsilon_{ji}$, subject to the Kuhn-Tucker condition $\mu_i(\lambda_i - y_{il} + \sum_{j \in \mathcal{L}_i \setminus l} y_{ji}) = 0$, with the Lagrange multiplier $\mu_i \leq 0 \; \forall i$. One can then show that [133]

$$\mu_i = \min \left[ 0, \left( \lambda_i - y_{il} + \sum_{j \in \mathcal{L}_i \setminus l} \left( \tilde{y}_{ji} - \frac{a_{j \setminus i} + \tilde{y}_{ji}}{b_{j \setminus i} + 1} \right) \right) \left( \sum_{j \in \mathcal{L}_i \setminus l} \frac{1}{b_{j \setminus i} + 1} \right)^{-1} \right], \quad (97)$$

$$\epsilon_{ji} = -\frac{a_{j \setminus i} + \tilde{y}_{ji} + \mu_i}{b_{j \setminus i} + 1}, \quad (98)$$

are the optimal solutions. If we interpret $\tilde{y}_{ji}$ as the current from $j$ to $i$ in the pervious iteration of the algorithm, then $\tilde{y}_{ji} + \epsilon_{ji}$ corresponds to the updated optimal current. This gives rise to a *backward message* of the algorithm to be explained later. To proceed, one differentiates $L_{i\backslash l}$ with respect to $y_{il}$ which gives

$$a_{i\backslash l} = \frac{\partial L_{i\backslash l}}{\partial y_{il}} = -\mu_i \tag{99}$$

$$b_{i\backslash l} = \frac{\partial^2 L_{i\backslash l}}{\partial y_{il}^2} = -\Theta(-\mu_i)\left(\sum_{j\in\mathcal{L}_i\backslash l} \frac{1}{b_{j\backslash i}+1}\right)^{-1}, \tag{100}$$

corresponding to the updated values of $a_{i\backslash l}$ and $b_{i\backslash l}$ of Eq. (95). Equations (98)-(100) constitute the update rules of the message passing algorithm.

To obtain the optimal configuration of currents by the above algorithm, the procedure is as follows:

1. pick randomly a node $i$ and select one of its neighbors as ancestor $l$,
2. compute and send to $l$ the values of $a_{i\backslash l}$ and $b_{i\backslash l}$ based on the estimated drawn current $y_{il}$,
3. compute and send to all descendants the corresponding values of $\epsilon_{ji}$, update $y_{ji}$ by $y_{ji} \to y_{ji} + \epsilon_{ji}$,
4. repeat steps 1-3 until the messages converge (i.e., change below a predefined limit).

The messages $a$ and $b$ thus constitute the *forward message* to ancestors and $\epsilon$ is the *backward message* to descendants. Physically, $y = \tilde{y} + \epsilon$ corresponds to the new point at which the descendants should compute the derivatives $a$ and $b$ of their cavity energy function $E(y)$. As the above algorithm is derived from the cavity equations under a replica symmetry-like (RS) ansatz, the convergence of the algorithm is an indication of RS stability. We note that the algorithm is *distributive*, as global optimizer and full knowledge of adjacency matrix is not required for individual nodes.

An alternative distributed algorithm can be derived by minimizing the Lagrangian

$$L = \sum_i \mu_i\left(\lambda_i + \sum_{j\in\mathcal{L}_i} y_{ji}\right) + \sum_{ji} \frac{y_{ji}^2}{2} \tag{101}$$

with respect to $y_{ji}$, where $\mu_i(\lambda_i + \sum_{j\in\mathcal{L}_i} y_{ji}) = 0$ is the Kuhn-Tucker condition with Lagrange multiplier $\mu_i \leq 0\ \forall i$. The optimal solution is given by

$$y_{ji} = \mu_j - \mu_i \tag{102}$$

$$\mu_i = \min\left[0, \frac{1}{k_i}\left(\lambda_i + \sum_{j\in\mathcal{L}_i} \mu_j\right)\right]. \tag{103}$$

Equation (103) is iterated until the convergence of all $\mu_i$, such that optimal configuration of currents is obtained by Eq. (102). If one considers $\mu_j$ and $\mu_i$ to be the potential of node $j$ and $i$, $y_{ji}$ can be interpreted as the potential difference between $j$ and $i$, analogous to electric circuits with uniform resistance.

Both algorithms are shown to converge in simulations and yield consistent results with theoretical predictions in cavity approach [133, 134]. The convergence of message passing algorithm indicates that the RS ansatz is sufficient to describe the phase space of the load balancing problem. It was shown in [134] that large currents are found around particularly rich and poor nodes (i.e. $\lambda \gg \langle\lambda\rangle$ and $\lambda \ll \langle\lambda\rangle$), as they provide and receive resources from others to achieve global satisfaction. Nodes with intermediate $\lambda$ mainly act as relays between the two groups and are surrounded by intermediate currents. A variant of the load balancing problem considers edges limited by bandwidth, in analogy to most networking systems. Behaviors such as bottleneck effect and clustering of balanced nodes are observed due to the restriction by edges bandwidth [137, 138].

### 4.2.2. The Optimal Location of Sources

Here we briefly describe the relation between the source allocation problem and decisions on the location of sources in networking systems. Unlike load balancing, the coefficient $\alpha$ in Eq. (84) is finite, which implies that nodes are allowed to be short of resources after allocation, i.e. $\lambda_i + \sum_{j \in \mathcal{L}_i} y_{ji} < 0$, when transportation cost is high. As $\psi(r)$ takes the form of a heaviside step function $\Theta(-r)$, the shortage cost is independent of the level of insufficient resources and the shortage nodes can be considered as a requirement for an installation of new resource providers. If one considers $\alpha$ to be the installation cost of a source node, e.g. transmitter or server, minimizing $H$ in Eq. (84) is equivalent to locating the optimal position for source nodes in the network [136].

One expects that when the installation cost $\alpha$ is small compared to transportation cost, source nodes are installed everywhere. When $\alpha$ increases, source nodes are installed only at optimal locations. It was shown [135, 136] that the system undergoes abrupt transitions when $\alpha$ increases, corresponding to the stability of resource penetration to increasing distance from the source nodes. In addition, it has been shown [135] that the source location problem is well described by the replica symmetry breaking (RSB) ansatz, and that the message passing algorithm derived from RS cavity equations does not converge. Algorithmic procedures such as decimation [129] have to be applied to obtain a good approximation of the optimal configuration [136].

### 4.2.3. Routing to Base Stations

Another variant of the resource allocation problem is relevant to optimizing path from individual sensors to a base station as shown in Fig. 8 [46]. In this case, one considers establishing a route from nodes to a base station as a hard constraint $\alpha \to \infty$ in the Hamiltonian $\mathcal{H}$ described by Eqs. (84) - (86). To achieve the task of path optimization as shown in Fig. 8, one makes two modifications to the model: (i) a specific distribution $\rho(\lambda)$ of initial resources, namely $\lambda = \infty$ for the base station, $\lambda = -1$ for transmitting sensor nodes and $\lambda = 0$ for all the other nodes; (ii) limit all the current variables $y_{ij}$ to be integers. Since each sensor is required to achieve non-negative final resources, the ground state of $\mathcal{H}$ is equivalent to an optimal configuration of paths from sensors to base stations, where path overlap (congestion) is suppressed via the introduction of a penalizing quadratic cost $\phi(y) \propto y^2$. This problem is also relevant to the Steiner tree problem which we will describe in Sec. 4.3.2, where shortest path between sensors and base stations are found without the consideration for congestion.

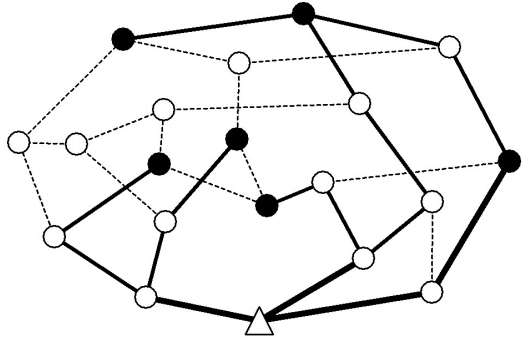

Figure 8: An example which show individual sensors (black circles) establishing a single path to a base station (triangle). Dashed lines correspond to idle links, while thin and thick solid lines correspond to communication loads of 1 and 2 units, respectively.

Compared to the case with continuous current, the cavity vertex energy $E^V_{j\setminus i}(y_{ji})$ in the present case has an integer domain and is no longer valid for Taylor series expansion used in the resource allocation problem (95) making the simplification by the piecewise quadratic function as in Eqs. (99) and (100) no longer applicable. Nevertheless, other simplifications can be made such that macroscopic properties of the optimal path configuration and an practical algorithm are derived [46].

### 4.3. Route optimization

In addition to frequency and resource allocation, we will review studies of disordered systems which are relevant to path optimization in networks. In particular, we will review the statistics of loops in networks, the minimum Steiner trees and a system of interacting polymers where the latter is relevant to routing between source-destination pairs in sparse networks.

### 4.3.1. Circuits and Loops in Networks

Circuits in networks are loops on the network topology, each of which taken on its own is devoid of intersections, i.e. is a self-avoiding closed path. Circuits are highly relevant to routing; for instance, finding a path which visits specific nodes on a graph with the lowest cost is termed the vehicle routing problem [139], which is relevant to the logistics of good delivery to multiple consumers by a single vehicle. Another related routing problem is the traveling salesman problem (TSP) [140], which involve finding the so-called Hamiltonian cycles which visit all the nodes exactly once. The number of cycles present in a graph serves as useful information for path planning in logistics problems. Marinari et al [141, 142] showed that the typical entropy of circuits with various length in graphs sampled from a given set of ensemble can be obtained using tools of statistical physics, in particular the quenched average entropy can be obtained by the cavity equations. Here we will review how such entropy can be obtained by following the line of Ref. [141].

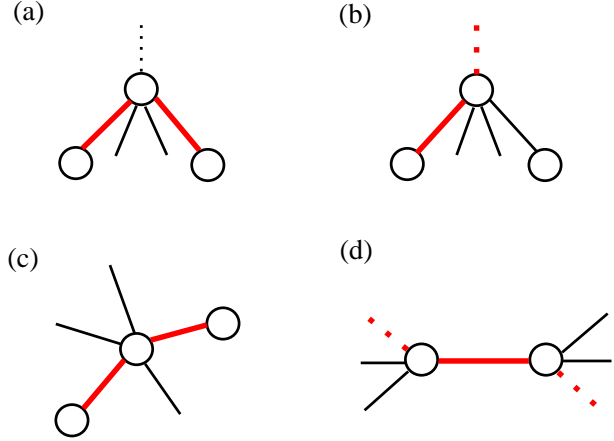

Figure 9: Schematic diagrams showing different alignment of the circuit, as represented by the thick red line.

To obtain the entropy of circuits, one defines a variable $s_{(ij)} = 1$ when the circuit passes the edge between $i$ and $j$ and otherwise $s_{(ij)} = 0$. One then considers the hamiltonian $\mathcal{H} = \sum_{(ij)} s_{(ij)}$, such that the partition function is given by [141]

$$Z(u) = \operatorname*{\mathbf{Tr}}_{\{s_{(ij)}\}} u^{\sum_{(ij)} s_{(ij)}} = \sum_L \mathcal{N}_L(\mathcal{A}) e^{-\beta L} \tag{104}$$

where $u = e^{-\beta}$, $L$ is the length of the circuit and $\mathcal{A}$ is the adjacency matrix. One further defines $f(u) = (1/N) \ln Z(u)$ and $\sigma(l) = (1/N) \ln \mathcal{N}_L$ with $l = L/N$, such that $Z(u) = \exp[N(l(u) \ln u + \sigma(l))]$. In the thermodynamic limit $N \to \infty$, the entropy $\sigma(l)$ of circuits with length $l$ is given by

$$\sigma(l) = f(u) - l(u) \ln u. \tag{105}$$

By cavity approach or free energy approximation which we will describe in Sec. 5.1.2, $f(u)$ and $l(u)$ are given by [141]

$$f(u) = \sum_{k=2}^{\infty} p(k) \int \prod_{j=1}^{k} [dy_j P(y_j)] \ln\left(1 + u^2 \sum_{i<j} y_i y_j\right)$$
$$- \frac{\langle k \rangle}{2} \int dy_i dy_j P(y_i) P(y_j) \ln(1 + u y_i y_j), \tag{106}$$

$$l(u) = \frac{\langle k \rangle}{2} \int dy_i dy_j P(y_i) P(y_j) \frac{u y_i y_j}{1 + u y_i y_j}. \tag{107}$$

where $p(k)$ is the degree distribution and the variable $y_i$ is proportional to the partition function of a case where the circuit passes node $i$ en route to the next node higher up the tree as shown in Fig. 9(b). In this case the terms $u^2 y_i y_j$ and $u y_i y_j$ in expressions (106) and (107) are proportional to the partition function of the circuit configuration as shown in Fig. 9(c) and (d) respectively, such that each of the thick red edges contributes a factor $u$ to the product. The distribution $P(y)$ is given self-consistently by

$$P(y) = q(0)\delta(y) + \sum_{k=1}^{\infty} q(k) \int \prod_{j=1}^{k} [dy_j P(y_j)] \delta\left(y - \frac{u \sum_j y_j}{1 + u^2 \sum_{i<j} y_i y_j}\right), \tag{108}$$

where $q(k)$ is the excess degree distribution and the terms $u^2 y_i y_j$ and $u y_j$ are proportional to the partition function of the circuit configuration as shown in Fig. 9(a) and (b) respectively. This equation resembles the form of the cavity equations (77) and (91) in the previous examples. The two terms in $f(u)$ resemble the form of the node and link energies in Eqs. (80) and (81) of the graph coloring problem and Eqs. (92) and (93) in resource allocation task, but instead of energy the two terms in $f(u)$ correspond to the free energy contribution from both node and link. The expression of $l(u)$ is equivalent to the energy of a link.

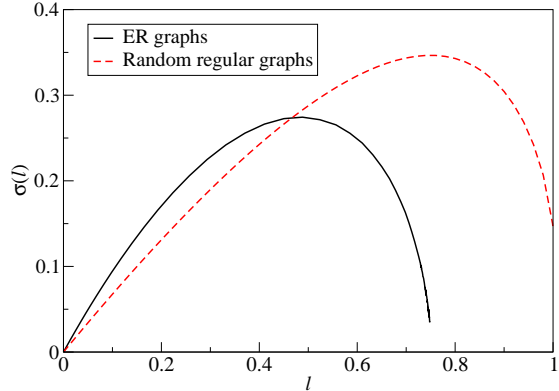

Figure 10: The entropy $\sigma(l)$ as a function of $l$ for ER graphs and random regular graphs, both with average degree $\langle k \rangle = 3$.

To evaluate the entropy of circuits, one varies $u$ in the range $0 < u < \infty$ and computes the corresponding distribution $P(y)$ from Eq. (108) and hence $f(u)$ and $l(u)$, and makes use of Eq. (105) to obtain $\sigma(l)$. Equation (108) can be solved by population dynamics, and the result of $\sigma(l)$ in ER graphs with average degree $\langle k \rangle = 3$ when $N \to \infty$ is shown in Fig. 10. We note that for regular graph with connectivity $k$, $P(y) = \delta[y - Y(k)]$ where $Y(k)$ is given by [141]

$$Y(k) = \sqrt{\frac{2u(k-1) - 2}{u^2(k-1)(k-2)}} \tag{109}$$

and can be easily obtained from Eq. (108). The $\sigma(l)$ of regular graph of connectivity $k$ is given by

$$\sigma(l) = -(1-l)\ln(1-l) + l\ln(k-1) + \left(\frac{k}{2} - l\right)\ln\left(1 - \frac{2l}{k}\right), \tag{110}$$

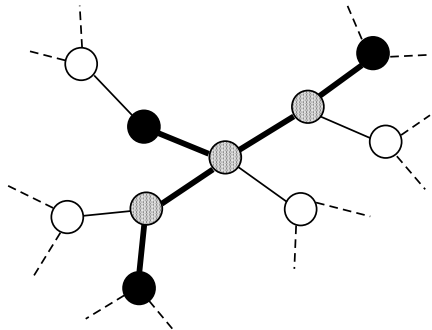

Figure 11: An example of Steiner tree in a simple network. The chosen of terminal nodes are colored in black, the Steiner nodes are colored in grey and the Steiner tree is represented by the bold line.

where the case of $k = 3$ is also shown in Fig. 10, exhibiting the existence of much higher probability of longer loops in comparison to the corresponding ER case.

Other than the entropy of circuits which was solved by the cavity approach, the traveling salesman problem (TSP) was also mapped into disordered systems and was studied by both replica [143] and cavity approaches [144]. Other studies of loops in networking systems include the empirical studies of short loops in the Internet at the level among autonomous systems (AS) [145], the loop statistics in finite size complex networks [146] and the derivation of algorithms to detect long loops in real instances [141, 147].

*4.3.2. The Minimum Steiner Trees*

The minimum spanning tree (MST) problem is extensively studied in mathematics, physics and computer science [148, 149]. Given an undirected network, a *spanning tree* of the network is a tree-like subgraph which connects the set of nodes. When edges are weighted, the MST decision problem concerns the existence of spanning trees with the sum of weight on edges being less than a given threshold. Finding the spanning tree with minimal edge weight thus serves as an optimization version of the MST problem. The search for MST in networks is related to the optimization of signal broadcast. Optimal broadcasting paths are defined as the spanning trees of minimal distance or traffic load on paths. The MST problem is also relevant to other networking systems such as transportation [150] and resistor networks [151].

A similar problem, known as the Steiner tree problem, is also relevant to routing and networking systems. Given an undirected network and a set of nodes, the *minimum Steiner tree* is defined as a tree of minimal weight which connects a given set of nodes, possibly with the help of other nodes. An example of a Steiner tree is given in Fig. 11, with the set of node to be connected colored in black and the minimum Steiner tree represented by the thicker edges. As one can see, the chosen nodes are usually the leaves of the tree and hence they are called *terminal nodes*. The non-terminal nodes which constitute the minimum Steiner tree are called *Steiner nodes*, and are shaded in grey in Fig. 11. Finding the minimum Steiner tree thus involves an optimal structure of the tree as well as an optimal set of Steiner nodes, which makes the related decision task an NP-complete problem [152].

While spanning trees are relevant to signal broadcast, Steiner trees are related to multicast communication with fixed recipients. The Steiner tree problem is thus similar to the source location problem [136] in the sense that optimal paths to a subset of nodes are found, but the recipients in the latter case are not fixed and installation of extra sources is required on the non-recipient nodes. One can also map Steiner trees to the optimal set of routers required to maintain the communication for in-use computers in a network. This is crucial for the energy-hungry Internet, as redundant routers can be switched off to save energy. The Steiner trees are also relevant to resilience of networking systems against router failures.

As compared to the graph coloring and resource allocation problems, the MST and Steiner tree problems involve global constraints which make analyses difficult. Nevertheless, one can adopt the cavity approach to derive a local message passing algorithm for obtaining the MST and Steiner tree solutions on networks.

To apply the cavity approach to the MST and Steiner tree problem, one can follow [153, 154] to derive the cavity equations for a general case of Steiner trees with depth of $D$ steps. In this case, one of the terminal node is chosen to be the *root* node, and the distances between the root node and all other terminal nodes $i$ in the Steiner trees are restricted to be at most $D$. One then defines the variable $p_i$ to be the *parent* of $i$ in the Steiner tree, such that $p_i = l$ with $l = 1, \cdots, N$ if $i$ is a terminal node, and $p_i = l$ or $p_i = \varnothing$ for non-terminal nodes. By setting all nodes to be terminal nodes, one obtains the MST. After defining the local variables, one can force the global constraints by noting that when $p_i = l$, $p_l \neq \varnothing$ and $d_i = d_l + 1$. These constitute constraints for the edge $(i, l)$ by which we define a variable $f_{il} = 1$ when they are satisfied and $f_{il} = 0$ otherwise. The variable $f_{il}$ can be expressed in terms of $f_{il} = g_{il} g_{li}$, with $g_{il}$ given by

$$g_{il} = [1 - \delta_{p_i,l}(1 - \delta_{d_i, d_l+1})](1 - \delta_{p_i,l}\delta_{p_l,\varnothing}). \qquad (111)$$

Similar to the cases of graph coloring and resource allocation, one can write a recursive relation of the cavity energy function $E_{i \setminus l}(p_i, d_i)$ as

$$E_{i \setminus l}(p_i, d_i) = w_{ip_i} + \sum_{j \in \mathcal{L}_i \setminus l} \min_{\{p_j, d_j | f_{ji}(p_j, d_j, p_i, d_i) = 1\}} [E_{j \setminus i}(p_j, d_j)], \qquad (112)$$

where $w_{ip_i} \geq 0$ is the weight on the edge $(i, p_i)$ and $w_{i\varnothing} = \infty$. The right hand side of Eq. (112) depends on $p_i$ and $d_i$ via the constraints $f_{ji}$. This recursive equation can be simplified by parametrization of $E_{i \setminus l}(p_i, d_i)$ in terms of cavity fields, as in the case of Eq. (72) in the graph coloring problem and Eq. (95) in the resource allocation problem. We refer the reader to [153, 154] for the derivation of the cavity fields in the case of Eq. (112), which leads to a message passing algorithm for solving the MST and Steiner trees problems on networks.

Other than via the cavity approach, the MST problem has also been analyzed by mapping it onto a disordered spin glass on regular lattices [155]. The scaling of the energy change subject to small perturbations about the optimal spanning tree was studied in [156].

*4.3.3. Routing as an Interaction of Polymers*

Another area of disordered systems relevant to networking is the study of interacting polymers on networks. In this case, one can consider the two ends of a polymer to be the source and the destination, and the polymer allocation on a network to represent the path, which is similar to the case shown in Fig. 1, where specific colors correspond to the respective polymers. One can then introduce interaction between the polymers which have a meaningful interpretation for networking, for instance, a repulsion between overlapping polymers can balance the individual occupancy of nodes/edges on the network and mitigate congestion on hubs. These methods have been devised in the area of polymer science [157] and have been used previously to study the TSP [143, 158] and the loop spectrum in networks [142].

To capture such interaction, one can define a Hamiltonian $\mathcal{H} \propto \sum_j (I_j)^\gamma$ as in [159] where $I_j$ is the number of polymers passing through node $j$. When $\gamma > 1$, overlap of polymers is penalized such that traffic congestions are suppressed. When $\gamma < 1$, overlap of polymers is encouraged and traffic tends to consolidate on common paths, leaving more nodes idle and thus can be switched off to reduce energy consumption. When $\gamma = 1$, there is no interaction between polymers and each of them is routed through the shortest path. To obtain the ground state of the polymer systems, and hence the optimal configuration of paths, Yeung et al [159] make use of the 0-vector method invented by de Gennes et al [157] which describes self-avoiding paths on network. As the replica calculation for solving the system is rather involved, we refer readers to Ref. [159] for details. Nevertheless, a non-monotonic trend of path length as a function of the number of polymers and a phase transition at $\gamma = 1$ which resembles the transition of the flow pattern of electric currents in resistor networks [150] is observed.

## 5. Probabilistic Inference for Networking

Inferring the state of variables $\widehat{x}$ by comparing their probability to be in the various states given a set of observations $\vec{z}$ is termed probabilistic inference. Accurate probabilistic inference can take the form of likelihood maximization, Maximum A Posteriori (MAP) and Marginal Posterior Maximization (MPM). The former merely requires a definition of the likelihood function for the given noise model while MAP and MPM require a prior assumption about the state of the variable; the difference between the two is that MAP examines the probability for the vector state $\vec{x}$ as a whole while MPM considers single variable marginals. The three methods can be summarized as:

$$\begin{aligned}
\widehat{x} &= \mathrm{argmax}_{\vec{x}}\ p(\vec{z}\mid\vec{x}) - \text{Maximum Likelihood} \\
\widehat{x} &= \mathrm{argmax}_{\vec{x}}\ p(\vec{x}\mid\vec{z}) - \text{MAP} \\
\widehat{x}_j &= \mathrm{argmax}_{x_j}\ p(x_j\mid\vec{z}) - \text{MPM}.
\end{aligned} \quad (113)$$

The three methods optimize different cost measures and give rise to different solutions [160, 161]. However, all three inference techniques are NP-hard and are computationally infeasible; one therefore resorts to principled approximation methods that are computationally efficient.

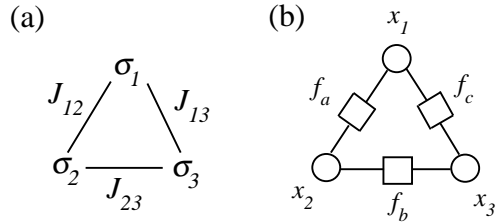

Figure 12: (a) A three-spin systems and (b) its corresponding factor graph representation.

There are several inference methods which provide efficient and principled estimation of marginal probabilities for individual variables in networks. For instance, belief propagation [161, 162] and variational approaches [163]. Here we are interested in belief propagation and its variants, leading to distributive algorithms with computational complexity that scales favorably with the system size, which are potentially applicable for large networking systems. They play an essential role in providing principled probabilistic inference in a broad range of applications from medical expert systems, to telecommunication. These methods, that have largely been developed independently in the computer science and information theory literature, also have deep roots in advanced mean field methods of statistical physics [161].

*5.1. Belief Propagation*

Here we describe the belief propagation (BP) algorithm that has been developed independently in different communities. The version introduced by Gallager [164] was applied to decoding in the context of error-correcting codes, Pearl [162] applied it to hierarchical Bayesian networks for estimating marginal probabilities and Mézard [37] introduced a macroscopic version of it for analyzing disordered systems. Links between the different frameworks were identified [161, 165] and extended to include more advanced approaches [166] that have been studied within the statistical physics community [167].

Belief propagation methods have been highly successful in a number of areas such as error-correcting codes [168], communication networks of sensors [169], calibration systems for sensor location [170], compressed sensing [171] and problems in statistical physic [39, 126]. The method was also applied to the study of combinational problems [125, 126, 129] and free energy approximation of physical systems.

By denoting $x_i$ as the state of node $i$, the goal of the BP algorithm is to compute the approximate marginal probability $p(x_i)$ for all $i$ in the network. The marginal probability $p(x_i)$ is given by $p(x_i) = \sum_{\vec{x}\backslash x_i} p(\vec{x})$, where the sum runs over all possible states of the system given that node $i$ is in state $x_i$. BP is usually formulated on bipartite *factor graphs* which consist of both *factor* and *variable* nodes. Regardless of network topology, many combinatorial problems, optimization problems and spin systems can be mapped onto a factor graph. We take the three spin system in Fig. 12(a) as an example. It can be mapped to a factor graph in Fig. 12(b) by introducing the factor nodes $f_a$, $f_b$ and $f_c$ on the edges, which characterize the interactions between the connected variable nodes. To see this, one can use an analogy with statistical physics and define

$$f_a(x_1, \cdots, x_k) = e^{-\beta E_a(x_1, \cdots, x_k)}, \tag{114}$$

where $E_a(x_1, \cdots, x_k)$ is the energy of the variables $x_1, \cdots, x_k$ through the factor $a$. For example, $f_a(x_1, x_2) = e^{-\beta J_{12} x_1 x_2}$ in the spin system of Fig. 12(a) and $f_a(x_1, x_2) = e^{-\beta \delta_{x_1, x_2}}$ for the graph coloring problem, where variables $x$ represent the assigned colors to the respective nodes. In this case, the probability that the system is in state $\vec{x}$ is given by [166]

$$p(\vec{x}) = \frac{1}{Z} e^{-\beta \sum_{a=1}^{M} E_a(\vec{x}_a)} = \frac{1}{Z} \prod_{a=1}^{M} f_a(\vec{x}_a), \tag{115}$$

where $M$ is the total number of factor nodes and $\vec{x}_a$ the state of the variable nodes connected to $a$.

*5.1.1. The Update Rules and Beliefs Computation*

To obtain an approximate pseudo-posterior for the marginal probability one assumes a local dependence of node and factor probabilities on their immediate neighborhoods. This assumption is exact on tree-like structures and provides a good approximation in many other systems. There are two types of messages in the BP algorithm. The first type of messages are from a factor node $i$ to a variable node $a$, which we denote by $u_{a \to i}(x_i)$; it corresponds to the conditional probability of factor $f_a$ given a node value $x_i$ - $p(f_a|x_i)$ up to normalization. The second type are messages from the variable node $i$ to factor node $a$, which we denote as $h_{i \to a}(x_i)$; it corresponds to the conditional probability of node $i$ taking the value $x_i$ given all the factors connected to it except for $f_a$ - $p(x_i|\{f\}_{\backslash a})$ - up to normalization. Update rules are obtained via simple Bayesian manipulations; to simplify the equations one writes the message $h_{i \to a}(x_i)$ as

$$h_{i \to a}(x_i) \propto \prod_{b \in \mathcal{L}_i \backslash a} u_{b \to i}(x_i), \tag{116}$$

and the message $u_{a \to i}(x_i)$ as

$$u_{a \to i}(x_i) = \sum_{\vec{x}_a \backslash x_i} f_a(\vec{x}_a) \prod_{j \in \mathcal{L}_a \backslash i} h_{j \to a}(x_j), \tag{117}$$

where the sum runs over all the possible states of variables in factor node $a$ given $i$ is in the state $x_i$. BP is also known as the *sum-product* algorithm as Eq. (117) involves a sum over products, and the BP messages are sometimes called *beliefs*. The messages can be initialized as appropriate for the problem at hand. We note that it is not necessary to normalize the messages as the normalization can be done when marginals of variables are computed. To compute the marginal of each variable node, we iterate the above update rules until convergence of all messages and evaluate the pseudo-posterior:

$$b(x_i) = \frac{1}{Z_i} \prod_{a \in \mathcal{L}_i} u_{a \to i}(x_i), \tag{118}$$

where $Z_i = \sum_{x_i} \prod_{a \in \mathcal{L}_i} u_{a \to i}(x_i)$ is a normalization constant, representing the approximate marginal probability $p(x_i)$ obtained by the BP algorithm

One notes that the products over messages $u_{b \to i}(x_i)$ in Eq. (116) and fields $h_{j \to a}$ in Eq. (117) implicitly assume that the corresponding conditional probabilities can be factorized, indicating statistical independence; BP is therefore exact only on tree networks. Nevertheless, it converges in many cases where the networks do include loops and hence is sometimes called the *loopy belief propagation* [172, 129, 169]. The convergence in loopy networks is quite remarkable; for instance, they converge to provide high quality solutions in a sensor network with asynchronous updating schedules, inhomogeneous communication rate on nodes and evolving datasets [169] as well as in many other applications.

The BP algorithm is equivalent to the cavity approach developed in the spin glass theory. One way to see this is to identify the messages $h_{i \to a}(x_i)$ with the partition function $Z_{i \to a}(x_i)$ of the network terminated at node $i$ of state $x_i$. One can combine Eqs. (116) and (117) to give

$$Z_{i \to a}(x_i) = \prod_{b \in \mathcal{L}_i \setminus a} \sum_{\vec{x}_b \setminus x_i} e^{-\beta E_b(\vec{x}_b)} \prod_{j \in \mathcal{L}_b \setminus i} Z_{j \to b}(x_j), \tag{119}$$

In cases where factor nodes connect only two variable nodes, such that $a$ connects $i$ to $l$, and $b$ connects $j$ to $i$, the above equation is rewritten as

$$Z_{i \to l}(x_i) = \sum_{\vec{x} \setminus x_i} e^{-\beta \sum_{j \in \mathcal{L}_i \setminus l} E_{ij}(x_i, x_j)} \prod_{j \in \mathcal{L}_i \setminus l} Z_{j \to i}(x_j), \tag{120}$$

where the sum runs over all the possible states $\vec{x}$ of the neighbors of $i$, given that $i$ is in state $x_i$. Finally, one introduces the free energy function $F_{i \setminus l}(x_i) = -T \ln Z_{i \to l}(x_i)$, $T$ being the temperature, such that in the zero-temperature limit, i.e. $\frac{1}{T} = \beta \to \infty$, the above expression reduces to the *min-sum* expression

$$F_{i \setminus l}(x_i) = \min_{\vec{x}} \left[ \sum_{j \in \mathcal{L}_i \setminus l} \left( F_{j \setminus i}(x_j) + E_{ij}(x_i, x_j) \right) \right], \tag{121}$$

which resembles Eq. (71) derived via the cavity approach. We note that the BP algorithm is analogous to the cavity approach with the replica symmetry ansatz, and is less likely to converge to optimal solutions in systems characterized by a multi-valley energy landscape. In this case, messages fickle between states in different macroscopically separated solution clusters and fail to converge. An improved version of BP called *Survey Propagation* (SP) [129] which considers a distribution of BP messages instead of a single BP message is described in Section 5.2.

*5.1.2. Free Energy Approximation*

In addition to its equivalence with the cavity approach, BP is useful for approximating the free energy of various systems. The simplest way is to make use of the Bethe approximation [173] and write the approximate free energy as $F = U - S$ with internal energy $U$ and entropy $S$ given by [166, 174]

$$U = -\sum_{a=1}^{M}\sum_{\vec{x}_a} b_a(\vec{x}_a) \ln f_a(\vec{x}_a), \tag{122}$$

$$S = -\sum_{a=1}^{M}\sum_{\vec{x}_a} b_a(\vec{x}_a) \ln b_a(\vec{x}_a) + \sum_{i=1}^{N}(k_i - 1)\sum_{x_i} b_i(x_i) \ln b_i(x_i). \tag{123}$$

The belief $b_a$ corresponds to the marginal probability of the factor node $a$, and the factor $k_i - 1$ in $S$ ensures the entropy of each variable node $i$ is counted exactly once. A variational approach can then be applied to find $b_i$ and $b_a$ that minimize the free energy $F$. Given that all factor nodes $a$ in the graph are characterized by positive $f_a(\vec{x}_a)$ for any $\vec{x}_a$, Yedidia et al [174, 166] showed that the local minima of $F$ are the BP fixed points with positive node and factor beliefs $b_i(= b(x_i))$ and $b_a$, respectively given by Eq. (118) and

$$b_a(\vec{x}_a) = \frac{1}{Z_a} \prod_{i \in \mathcal{L}_a} \prod_{b \in \mathcal{L}_i \setminus a} u_{b \to i}(x_i) \tag{124}$$

where $Z_a = \sum_{\vec{x}_a} \prod_{i \in \mathcal{L}_a} \prod_{b \in \mathcal{L}_i \setminus a} u_{b \to i}(x_i)$ is a normalization constant.

As loops are always present in networks in general and in structured networks in particular, Yedidia et al [166] generalized the BP algorithm to include the influence of loops in node clusters in a similar manner to the cluster variation approach [167]. The *generalized belief propagation* (GBP) algorithm considers factor graphs divided into overlapping regions, with local loops allowed in each regions. Various approaches are introduced to identify regions in a factor graph, and the accuracy of GBP greatly depends on the division of the network to regions. Once the regions are identified, messages are passed between regions instead of between single nodes. It is shown [166] that the GBP algorithm converges in cases where BP does not and results in a better estimate of the marginal probabilities; this is particularly true in structured networks, for instance in image restoration [175]. As more accurate marginal probabilities are computed by the GBP, it also gives a better approximation of the free energy [166]. Other approaches to address the problem of loops has been presented [176, 177] and rely on increasingly more complex correlations between network constituents.

*5.2. Survey Propagation*

Convergence of the BP algorithm has been observed to fail in systems characterized by numerous low-lying states separated by high energy barriers (see for instance Fig. 13); i.e., macroscopically separated solution clusters. One of the reasons for this failure is that individual BP messages tend to approach states in different valleys, i.e. different solution clusters, which leads to inconsistent messages. In view of the problem, Mézard and Zecchina developed an improved version of BP, known as the Survey Propagation (SP), which incorporates the structure of the energy landscape [129, 178] into the cavity equations (or equivalently the BP equations) through the introduction of a distribution of messages.

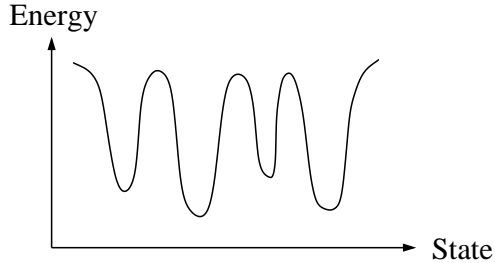

Figure 13: An energy landscape with multiple valleys.

One can see in examples such as the graph coloring [125, 126], $K$-satisfiability [129] and the resource allocation problem [135] that the analytical results derived from the SP cavity equations are equivalent to the those derived from the 1RSB replica approach.

To understand the underlying physical assumptions of SP, one can identify an energy valley as a *pure state* of the system. Each pure state is characterized by a particular set of BP messages or cavity fields. Systems with multiple valleys are thus described by a distribution of BP messages for each edge, instead of one particular BP message. One can further assume that the number of pure states is a function of average energy $\epsilon = E/N$, given by [129]

$$\mathcal{N}(\epsilon) = \exp[N\sigma(\epsilon)]. \tag{125}$$

The function $\sigma(\epsilon)$ is called the *complexity* of the system such that $N\sigma(\epsilon)$ plays the role of entropy among pure states. This is to be distinguished from the *internal entropy* within a pure state, where different degenerate configurations of on-site variables are considered.

*5.2.1. The Update Rules and Decimation Procedures*

To derive the SP cavity equation, one first expands the complexity about a reference energy $\epsilon_0$ as

$$\sigma(\epsilon) \approx \sigma(\epsilon_0) + x(\epsilon - \epsilon_0) \tag{126}$$

with $x = \partial\sigma/\partial\epsilon$. By fixing the value of $x$, one is effectively interested at pure states with average energy $\epsilon_0$, as well as states with small deviation from $\epsilon_0$. One then writes a recursion equation for the distribution $P_{i\backslash l}(h_{i\backslash l})$ of the cavity field at $i$ in the absence of $l$, as

$$P_{i\backslash l}(h_{i\backslash l})\mathcal{N}(\epsilon_{i\backslash l}) \propto \prod_{j\in\mathcal{L}_i}\left[\int dh_j P_{j\backslash i}(h_{j\backslash i})\mathcal{N}(\epsilon_{j\backslash i})\right]\delta[h_{i\backslash l} - \mathcal{H}(h_{j_1\backslash i}, \cdots, h_{j_{k_i}\backslash i})]. \tag{127}$$

The products $P\mathcal{N}$ on both sides of this equation preserve the average energy close to $\epsilon_0$ by a mechanism similar to detailed balance. For instance, if $\sigma(\epsilon)$ is increasing around $\epsilon_0$, states with small average energy $\epsilon$ (hence a small $\mathcal{N}$) are multiplied proportionally by a large $P$ to counterbalance states with large $\epsilon$ (which have a large $\mathcal{N}$). The distribution $P_{i\setminus l}(h_{i\setminus l})$ is called the *survey* of cavity field among pure states, which leads to the name of the algorithm-survey propagation [129]. The function $\mathcal{H}(h_{j_1\setminus i}, \cdots, h_{j_{k_i}\setminus i})$ in Eq. (127) corresponds to the recursion relation of cavity fields.

By using Eqs. (125) and (126), one can simplify Eq. (127) to obtain

$$P_{i\setminus l}(h_{i\setminus l}) \propto \prod_{j\in\mathcal{L}_i} \left[\int dh_j P_{j\setminus i}(h_{j\setminus i})\right] \delta[h_{i\setminus l} - \mathcal{H}(h_{j_1\setminus i}, \cdots, h_{j_{k_i}\setminus i})]e^{-x\Delta E}, \qquad (128)$$

where $\Delta E$ is the change of energy with the addition of node $i$ to the original network. In other words,

$$\Delta E = \min_{\sigma_i}[E_{i\setminus l}(\sigma_i)] - \sum_{j\in\mathcal{L}_i\setminus l} \min[E_{j\setminus i}(\sigma_j)], \qquad (129)$$

where $E_{i\setminus l}(\sigma_i)$ is the cavity energy function of the network terminated at $i$, and $\sigma_i$ the state of variable $i$. The term $e^{-x\Delta E}$ in Eq. (128) is a reweighing factor of the cavity states with different $\epsilon$ values, in order to keep the average energy close to $\epsilon_0$. With a particular choice of $x$, the above equation can be iterated on the network until $P_{j\setminus i}(h_{j\setminus i})$ converges, which corresponds to the distribution of cavity fields at $i$ among the pure states with energy density $\epsilon_0$.

For inference in specific instances, one runs the SP cavity equations together with a *decimation* process. The procedures are as follows. One first chooses a value of $x$ according the form of $\sigma(\epsilon)$, which corresponds to the $\epsilon$ value of interest. We refer the reader to [129] for computation of $\sigma(\epsilon)$ in specific instances. The cavity equations are then iterated on the network until convergence; nodes with the highest bias towards one of the states (as given by all neighboring $P_{j\setminus i}(h_{j\setminus i})$) are then *decimated*, i.e. their state become fixed to the state with the highest probability. The SP equations are then re-run, with all the decimated nodes fixed, until either all nodes are decimated or when the SP solution reduces to a BP solution, i.e. $P_{j\setminus i}(h_{j\setminus i}) = \delta(h_{j\setminus i} - h_{j\setminus i}^{\mathrm{BP}})$ for all remaining nodes. This final state corresponds to the solution obtained by the SP algorithm. In some cases another algorithm such as WalkSat is used at the final stage of the process [179].

*5.3. Network Tomography and Compressed Sensing*

We have focused so far on the use of distributed inference methods for optimization and on identifying the most probable state of individual network nodes, there is another important task relevant to networking which falls within the field of statistical interference - network tomography.

Network tomography corresponds to the recovery of the internal characteristics of a network from the end-point data. A typical case of network tomography that is relevant to networking is based on monitoring link metrics, such as delay or packet loss, in communication networks through a small number of end-to-end data [180]. For instance, if we denote the delay of an individual link $l$ as $x_l$, the end-to-end delay along path $p$ as $y_p$, the number of links as $M$ and the number of path on which performance are measured as $P$, then the $M$-component vector $\vec{x}$ and $P$-component $\vec{y}$ are related by

$$\vec{y} = \mathcal{A} \cdot \vec{x}. \tag{130}$$

where $\mathcal{A}$ is a $P \times M$ matrix such that $a_{pl} = 1$ if path $p$ passes through the edge $l$, and $a_{pl} = 0$ otherwise. Physically, Eq. (130) assumes that the delay along a path $p$ is the sum of the delays incurred on all of its constituent links. To obtain the delay on each of the $M$ individual links, one can measure delays along $P$ paths to obtain the vector $\vec{y}$ and solve the system of linear equations described in Eq. (130) for $\vec{x}$.

To find all the components of $\vec{x}$ accurately, one normally requires at least $P = M$ measurements. However, if $\vec{x}$ is a sparse vector, i.e. only a fraction of links have non-zero delays, one can employ techniques used in *compressed sensing* [36] to solve the systems by making $P < M$ measurements. In this case of sparse $\vec{x}$, one can achieve the goal by minimizing the $L_0$ norm of the a vector $\vec{x}_p$ [36], i.e. the number of non-zero element in $\vec{x}_p$, subject to the constraint $\vec{y} = \mathcal{A} \cdot \vec{x}_p$ given by Eq. (130). Minimization of the $L_0$ norm is known to saturate the *theoretical* limit $\alpha = \rho$ shown by the dashed line in Fig. 14; however, as minimization of the $L_0$ norm is typically computationally difficult and most studies focus on $L_1$ norm minimization instead i.e. finding the vector $\vec{x}_s$ such that

$$\vec{x}_s = \operatorname*{argmin}_{\vec{x}_p} ||\vec{x}_p|| \tag{131}$$

subject to the constraint $\vec{y} = \mathcal{A} \cdot \vec{x}_p$ given by Eq. (130). One thus expects the above optimization problem leads to a vector $\vec{x}_s$ that coincide with the true state $\vec{x}$. The formulation of the problem allows for the use of statistical physics approaches, such as the replica and cavity methods, for carrying out analyses of typical or macroscopic properties. If we denote $\alpha = P/M$ and the fraction of non-zero elements in $\vec{x}$ as $\rho$, Kabashima et al [181] and Ganguli et al [182] have shown that the reconstruction always leads to the correct state $\vec{x}$ when $\alpha > \alpha_c(\rho)$ in the limit $M \to \infty$. While for $\alpha < \alpha_c(\rho)$ the probability for finding the correct solution tends to zero as $M \to \infty$. Hence the critical value $\alpha_c(\rho)$ marks the transition between the two phases, and is given by the solution of the following equations, by eliminating the variable $x$ [181, 182]

$$\alpha = 2(1-\rho)\left[(1+x^2)H(x) - x\frac{e^{-x^2/2}}{\sqrt{2\pi}}\right] + \rho(1+x^2) \tag{132}$$

$$\alpha = 2(1-\rho)H(x) + \rho \tag{133}$$

where $H(x) \equiv \frac{1}{\sqrt{2\pi}} \int_x^\infty dz e^{-z^2/2}$, i.e. a standard Gaussian integral. The results of $\alpha_c(\rho)$, and thus the phase diagram for compressed sensing obtained by minimizing the $L_1$ norm, are shown in Fig. 14. We refer readers to Refs [181, 182] for the derivations of the above equations by replica approach.

Recent analysis by Krzakala et al [183] showed the potential of BP based methods in going beyond the $L_1$ theoretical limit $\alpha_c(\rho)$ under various sparsity conditions and assumptions and suggest a clever design of the sampled data that leads to a crystallization effect in the inference dynamics and saturates the $L_0$ theoretical limit $\alpha = \rho$. This is conditioned on the ability to design the obtained measurements and is not applicable for random measurements.

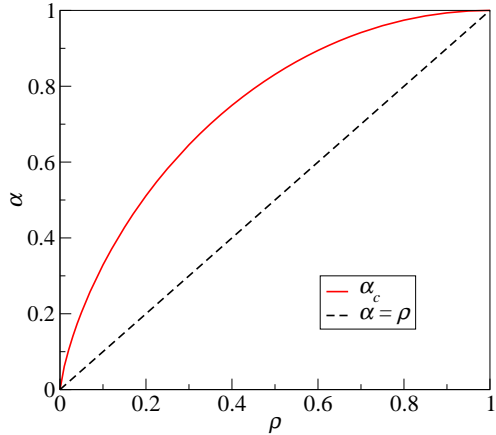

Figure 14: The phase diagram for compressed sensing by minimizing the $L_1$ norm, obtained in [181, 182]. Correct reconstruction is achieved in the region with $\alpha > \alpha_c(\rho)$. The dashed line represents the theoretical limit $\alpha = \rho$ obtained via $L_0$ norm minimization.

In addition to the tomography of link delays in communication networks, compressed sensing can be applied to detect attenuation loss in a region for surveillance purposes [184], or to detect the location of moving sensors [185]. Another area of research in physics which is related to the tomography of networking from dynamical data are studies of the inverse inference problem in Ising models, whereby one is given data of dynamically changing local magnetizations and correlations between spins, to infer the couplings between all node pairs and the local magnetic field at each node [186, 187]. Due to the compatibility of these models with a dynamically evolving network data, they may be applied to infer the underlying topology in networking systems.

## 6. Future Directions and Discussions

We showed that methodology from statistical physics is readily available to be used for modeling and analysis of networking systems. Directly applicable outcomes mostly include practical algorithms, for example, searching and routing algorithms from preferential random walk and the cavity and message passing algorithms, which can be used to minimize frequency interference and transportation cost. In addition to practical algorithms; and equally importantly, statistical physics offers a macroscopic overview which predicts system's behavior subject to the choice of parameters. Examples include the evaluation of epidemic thresholds for various network types, the sensitivity of networks to cascading failures, the prediction of phase transitions to algorithmically difficult regimes in hard combinatorial problems and the distribution of resources under different topologies and initial resource distributions. These theoretical macroscopic findings allow decision makers to foresee potential problems and mitigate them through better design and by taking in advance precautionary steps. Such macroscopic insights and forecasts establish an essential role for statistical physics in the study of networking systems.

We remark that the present review aims mainly to address the most essential topics at the interface between physics and networking. There exist several other areas in statistical physics relevant to networking including the synchronization in oscillator networks [20], critical phenomenon in complex networks [18], scaling properties of the Internet [56, 188, 189] and other communication systems. We refer the reader to these references for details.

Although physicists have already made significant progress in the study of networking systems, more consolidated efforts are required and expected. We note that the models studied so far are simple and capture only the fundamental aspects of networking and are far from capturing all properties of real networking systems. On the one hand, simple models facilitate theoretical analyses and make the macroscopic picture more coherent and clear; on the other hand, some emergent features may be lost in the simplified picture. To have a comprehensive understanding of networking, it is thus crucial to build on the existing framework, which preserves the compatibility with statistical physical techniques, a more realistic model of networking systems. For instance, one may consider models with multi-path communication by file segmentation, buffers of limited capacity in routing and specific constraints of peer-to-peer networks. We expect to see such progress and their applications in the near future.

## Acknowledgements

Support by the EU FET FP7 project STAMINA (FP7-265496) and the Royal Society Exchange Grant IE110151 is acknowledged. We would like to thank Alexander Stepanenko and Marc Mézard for helpful comments on the manuscript.